\documentclass[reprint,superscriptaddress,nofootinbib,amsmath,amssymb,aps,prx]{revtex4-2}

\usepackage{silence}
\usepackage{bm,graphicx,float,url,hyperref,cleveref,braket,qcircuit,mathtools,amsthm}

\crefname{figure}{Fig.}{Figs.}
\crefname{equation}{Eq.}{Eqs.}
\crefname{section}{Sec.}{Sec.}
\crefname{table}{Table}{Tables}
\Crefname{figure}{Figure}{Figures}
\Crefname{equation}{Equation}{Equations}
\Crefname{section}{Section}{Sections}
\Crefname{table}{Table}{Tables}

\theoremstyle{definition}

\begin{document}

\preprint{APS/123-QED}
\title{Systematic construction of multi-controlled Pauli gate decompositions with optimal \texorpdfstring{$T$}{T}-count}

\author{Ken M. Nakanishi}
    \email{ken-nakanishi@g.ecc.u-tokyo.ac.jp}
    \affiliation{
        Institute for Physics of Intelligence,
        The University of Tokyo,
        Tokyo 113-0033, Japan.
    }
\author{Synge Todo}
    \email{wistaria@phys.s.u-tokyo.ac.jp}
    \affiliation{
        Department of Physics,
        The University of Tokyo,
        Tokyo 113-0033, Japan.
    }
    \affiliation{
        Institute for Physics of Intelligence,
        The University of Tokyo,
        Tokyo 113-0033, Japan.
    }
    \affiliation{
        Institute for Solid State Physics,
        The University of Tokyo,
        Kashiwa, 277-8581, Japan.
    }

\date{\today}

\begin{abstract}
Multi-controlled Pauli gates are typical high-level qubit operations that appear in the quantum circuits of various quantum algorithms. We find multi-controlled Pauli gate decompositions with smaller CNOT-count or $T$-depth while keeping the currently known minimum $T$-count. For example, for the CCCZ gate, we find decompositions with CNOT-count 7 or $T$-depth 2 while keeping the $T$-count at the currently known minimum of 6. The discovery of these efficient decompositions improves the computational efficiency of many quantum algorithms. What led to this discovery is the systematic procedure for constructing multi-controlled Pauli gate decompositions. This procedure not only deepens our theoretical understanding of quantum gate decomposition but also leads to more efficient decompositions that have yet to be discovered.
\end{abstract}

\maketitle

\section{Introduction}

\begin{figure}[t]
\vspace{3.8ex}
\flushleft{(a)}\vspace{-6ex}
\begin{equation*}
\begin{array}{c}\scalebox{.8}{\Qcircuit @C=.3em @R=-.1em @!R {
&&\qw&\qw&\qw&\qw&\qw&\qw&\qw&\qw&\qw&\qw&\ctrl{4}&\qw&\qw&\qw&\qw&\qw&\ctrl{1}&\qw&\qw&\qw\\
&&\qw&\qw&\qw&\qw&\qw&\qw&\qw&\qw&\ctrl{3}&\qw&\qw&\qw&\ctrl{3}&\qw&\qw&\qw&\controlo\qw&\qw&\qw&\qw\\
&&\qw&\qw&\qw&\qw&\qw&\ctrl{2}&\qw&\qw&\qw&\qw&\qw&\qw&\qw&\qw&\qw&\qw&\qw&\ctrl{1}&\qw&\qw\\
&&\qw&\qw&\qw&\ctrl{1}&\qw&\qw&\qw&\ctrl{1}&\qw&\qw&\qw&\qw&\qw&\qw&\qw&\qw&\qw&\control\qw&\qw&\qw\\
&\push{\ket{0}}&&\gate{H}&\gate{T}&\targ&\gate{T^\dagger}&\targ&\gate{T}&\targ&\targ&\gate{T}&\targ&\gate{T^\dagger}&\targ&\gate{T}&\gate{H}&\meter&\cctrl{-3}&\cctrl{-1}&&
}}\end{array}
\end{equation*}
\flushleft{(a')}\vspace{-6ex}
\begin{equation*}
\begin{array}{c}\scalebox{.8}{\Qcircuit @C=.3em @R=-.1em @!R {
&&\qw&\qw&\qw&\qw&\qw&\qw&\qw&\qw&\qw&\qw&\qw&\ctrl{4}&\qw&\qw&\qw&\qw&\qw&\control\qw&\qw&\qw&\qw\\
&&\qw&\qw&\qw&\qw&\qw&\qw&\qw&\qw&\qw&\ctrl{3}&\qw&\qw&\qw&\ctrl{3}&\qw&\qw&\qw&\ctrlo{-1}&\qw&\qw&\qw\\
&&\qw&\qw&\qw&\qw&\qw&\ctrl{2}&\qw&\qw&\qw&\qw&\qw&\qw&\qw&\qw&\qw&\qw&\qw&\qw&\ctrl{1}&\qw&\qw\\
&&\qw&\qw&\qw&\ctrl{1}&\qw&\qw&\qw&\ctrl{1}&\qw&\qw&\qw&\qw&\qw&\qw&\qw&\qw&\qw&\qw&\control\qw&\qw&\qw\\
&\push{\ket{0}}&&\gate{H}&\gate{T}&\targ&\gate{T^\dagger}&\targ&\gate{T}&\targ&\gate{S^\dagger}&\targ&\gate{T^\dagger}&\targ&\gate{T}&\targ&\gate{T^\dagger}&\gate{H}&\meter&\cctrl{-3}&\cctrlo{-1}&&
}}\end{array}\smallskip
\end{equation*}
\flushleft{(b)} \vspace{-3.8ex}
\begin{equation*}
\begin{array}{c}\scalebox{.8}{\Qcircuit @C=.3em @R=.2em @!R {
&&\qw&\qw&\qw&\qw&\qw&\qw&\qw&\qw&\ctrl{4}&\targ&\gate{T^\dagger}&\targ&\qw&\qw&\ctrlo{1}&\qw&\qw&\qw\\
&&\qw&\qw&\qw&\qw&\qw&\qw&\qw&\ctrl{3}&\qw&\targ&\gate{T^\dagger}&\targ&\qw&\qw&\controlo\qw&\qw&\qw&\qw\\
&&\qw&\qw&\targ&\gate{T^\dagger}&\targ&\qw&\ctrl{2}&\qw&\qw&\qw&\qw&\qw&\qw&\qw&\qw&\ctrl{1}&\qw&\qw\\
&&\qw&\qw&\targ&\gate{T^\dagger}&\targ&\ctrl{1}&\qw&\qw&\qw&\qw&\qw&\qw&\qw&\qw&\qw&\control\qw&\qw&\qw\\
&\push{\ket{0}}&&\gate{H}&\ctrl{-2}&\gate{T}&\ctrl{-2}&\targ&\targ&\targ&\targ&\ctrl{-4}&\gate{T}&\ctrl{-4}&\gate{H}&\meter&\cctrl{-3}&\cctrl{-1}&&
}}\end{array}
\end{equation*}
\caption{CCCZ gate decompositions with $T$-count 6. (a) CCCZ decomposition with CNOT-count 6 or 8. (a') CCCZ decomposition with CNOT-count 7. (b) CCCZ decomposition with $T$-depth 2.}
\label{fig:comp_cccz}
\end{figure}

Quantum algorithms perform computations by combining operations on the states of qubits. To date, many groundbreaking algorithms have been proposed~\cite{shor1994algorithms,shor1999polynomial,grover1996fast,hallgren2007polynomial,hallgren2005fast,schmidt2005polynomial,van2002efficient,donis2017quantum,van2008classical,jordan2008fast,bernstein2013quantum,boneh1995quantum,ettinger2004quantum,hallgren2000normal,bernstein1993quantum,deutsch1985quantum,reichardt2011reflections,van2006quantum,bennett1997strengths,durr2006quantum,ambainis2011quantum,childs2003exponential,magniez2007quantum,itakura2005quantum,magniez2007quantum2,childs2007quantum,liu2009quantum,mosca1999quantum,cheung2001decomposing,bravyi2011quantum,arvind2006complexity,iwama2012quantum,belovs2011span,ambainis2007quantum,childs2003quantum,ambainis2012quantum,ambainis2006quantum,wang2013efficient,ambainis2016efficient,freedman2002simulation,freedman2002modular,alagic2010estimating}, including Shor's algorithm~\cite{shor1994algorithms,shor1999polynomial} and Grover's algorithm~\cite{grover1996fast}, which are expected to outperform classical computers in specific tasks. In addition, many algorithms have been proposed that make use of the currently available noisy quantum computers~\cite{Boixo2018,bouland2019complexity,chen2018classical} in combination with classical computers~\cite{Peruzzo2014,mcclean2020openfermion,Bauer2016,Kandala2017,nakanishi2019subspace,Heya2018,Farhi2014,Farhi2016,Otterbach2017,Biamonte2017,benedetti2019parameterized,Mitarai2018,Wiebe2014,havlivcek2019supervised,Lloyd2013,Khoshaman2018,Dallaire2018}. Quantum algorithms are described by quantum circuits, which are sequences of quantum gates that represent the operations on the state of qubits. To execute a quantum circuit on a quantum computer, it is necessary to perform quantum circuit decomposition, which transforms the high-level qubit operations contained in the quantum circuit into a sequence of low-level one- or two-qubit gates that can be executed directly on the quantum computer. We consider as low-level gates the $T$, $S$, Hadamard, and CNOT gates, as well as the $T^\dagger$, $S^\dagger$, $Z$, $X$, and CZ gates, which can be easily constructed using them.

In quantum gate decomposition, it is important to minimize the computational resources used by the decomposed quantum gates. In this paper, we consider the following three metrics for these computational resources:
\begin{description}
    \item[$T$-count] The number of $T$ and $T^\dagger$ gates used in a quantum circuit. In fault-tolerant quantum computing, $T$ and $T^\dagger$ gates consume significantly more resources than Clifford gates~\cite{bravyi2011quantum,fowler2012surface}. Therefore, it is crucial to minimize the number of $T$ and $T^\dagger$ gates used in a quantum circuit~\cite{selinger2013quantum,jones2013low,gidney2018halving,gidney2021cccz,paler2022realistic,beverland2020lower,jiang2023lower,de2020fast,Kliuchnikov2016,Ross2016,Gosset2014,Mosca2022,amy2014polynomial,gheorghiu2022t,maslov2016advantages,zindorf2024efficient,Brugiere2021}.
    \item[CNOT-count] The number of CNOT and CZ gates used in a quantum circuit. Since CNOT gates require more computational resources than single-qubit gates on current non-fault-tolerant quantum computers, it is important to minimize the number of CNOT gates used in a quantum circuit~\cite{zindorf2024efficient,Schuch2003-ze,Schuch_undated-yw,mottonen2004quantum,shende2009cnot,park2023reducing,Brugiere2021,maslov2016advantages,liu2023feasibility,liu2023minimizing,nemkov2023efficient,nakanishi2024decompositions}. Since a CZ gate can be converted into a CNOT gate using two Hadamard gates, we count one CZ gate as equivalent to one CNOT gate when considering the CNOT-count. Recently, a study was published showing that $T$ states, the resource for $T$ gates in fault-tolerant quantum computers, can be grown as cheaply as CNOT gates~\cite{gidney2024magic}. Therefore, minimizing the CNOT-count becomes just as important as minimizing the $T$-count in future research on fault-tolerant quantum computers as well as current non-fault-tolerant quantum computers.
    \item[$T$-depth] The number of $T$-stages in a quantum circuit~\cite{amy2013meet}. A $T$-stage is defined as a set of one or more $T$ or $T^\dagger$ gates that can be executed simultaneously on different qubits. $T$-depth represents the execution time of the quantum circuit in terms of the number of $T$ gates, ignoring the time required to execute Clifford gates~\cite{amy2013meet,selinger2013quantum,jones2013low,gidney2018halving,gidney2021cccz,paler2022realistic,paler2023t,zindorf2024efficient,amy2014polynomial,gheorghiu2022t,gheorghiu2022quasi,Niemann2019,lee2023t,huang2022synthesizing,maslov2016advantages,Brugiere2021}. Therefore, reducing $T$-depth is important for shortening the overall execution time.
\end{description}

As high-level quantum gates, we consider multi-controlled Pauli gates. There are three multi-controlled Pauli gates: multi-controlled NOT gate, multi-controlled $Y$ gate, and multi-controlled $Z$ gate. Multi-controlled Pauli gates are typical high-level qubit operations that appear in the quantum circuits of various quantum algorithms. For example, the oracle in Grover's algorithm can be implemented using a multi-controlled $Z$ gate and several $X$ gates. The multi-controlled NOT gate is a generalization of the CNOT and Toffoli gates. The $n$-controlled $Z$ gate is also written as the C$^n$Z gate, and when $n=1,2,3$, it is also written as the CZ, CCZ, and CCCZ gates, respectively. The three multi-controlled Pauli gates can be easily converted into each other using Hadamard gates and/or $S$ gates. There are many studies on the decomposition of multi-controlled Pauli gates~\cite{amy2013meet,selinger2013quantum,jones2013low,gidney2018halving,gidney2021cccz,paler2022realistic,paler2023t,zindorf2024efficient,beverland2020lower,nie2024quantum,Barenco1995-ro,nemkov2023efficient,saeedi2013linear,amy2014polynomial,shende2009cnot}. For example, Jones~\cite{jones2013low} found an $n$-controlled Pauli gate decomposition with $T$-count $4n-4$, and later Gidney and Jones~\cite{gidney2021cccz} improved this to $T$-count $4n-6$ when $n \geq 3$.

In this paper, we focus on the decompositions of multi-controlled Pauli gates when the low-level quantum gates are $T$ gate, CNOT gate, and typical one-qubit Clifford gates, considering both fault-tolerant and non-fault-tolerant quantum computation. We find a systematic procedure to construct multi-controlled Pauli gate decompositions with smaller CNOT-count or $T$-depth while keeping the currently known minimum $T$-count. For example, for the CCCZ gate, we find a decomposition with CNOT-count 7 or $T$-depth 2, while keeping the $T$-count at the currently known minimum of 6. Since multi-controlled Pauli gates are used in the quantum circuits of many quantum algorithms, the discovery of these efficient decompositions improves the computational efficiency of many quantum algorithms. This result is obtained through a systematic procedure for constructing multi-controlled Pauli gate decompositions. This procedure not only deepens our theoretical understanding of quantum circuit decompositions but also leads to more efficient gate decompositions that have yet to be found.

The rest of the present paper is organized as follows. In \cref{sec:previous_studies}, we summarize previous research on the Clifford+$T$ gate decomposition of multi-controlled Pauli gates. In \cref{sec:improve_selinger}, we show that the $T$-depth of the multi-controlled Pauli gate decomposition as described by Selinger~\cite{selinger2013quantum} can be improved a little more. In \cref{sec:main_decompositions}, we derive a systematic construction method for decompositions with smaller CNOT-count or $T$-depth while keeping the currently known minimum $T$-count. In \cref{sec:comp}, we compare our three novel CZ gate decompositions with previous research. \Cref{sec:conclusion} is devoted to the conclusion.

\section{Previous studies}\label{sec:previous_studies}
In this section, we review the previous research on the decompositions of multi-controlled Pauli gates. The three types of multi-controlled Pauli gates (multi-controlled NOT gate, multi-controlled $Y$ gate, and multi-controlled $Z$ gate) can be transformed into each other using Hadamard gates and/or $S$ gates as
\begin{equation}
\begin{array}{c}\scalebox{1}{\Qcircuit @C=.5em @R=0em @!R {
&{/}\qw&\ctrl{1}&\qw&\qw&&&{/}\qw&\ctrl{1}&\qw&\qw&&&{/}\qw&\ctrl{1}&\qw&\qw\\
&\qw&\targ&\qw&\qw&\push{\raisebox{1em}{=}}&&\gate{H}&\control\qw&\gate{H}&\qw&\push{\raisebox{1em}{=}}&&\gate{S}&\gate{Y}&\gate{S^\dagger}&\qw
}}\end{array}.\label{eq:cnx_cny_cnz}
\end{equation}
Thus, finding a decomposition with $T$-count $a$, CNOT-count $b$, and $T$-depth $c$ for any one of the three $n$-controlled Pauli gates is equivalent to finding a decomposition with $T$-count $a$, CNOT-count $b$, and $T$-depth $c$ for all three $n$-controlled Pauli gates.

The following quantum circuit represents a standard decomposition of the Toffoli gate:
\begin{equation}\label{eq:toffoli_standard_decomposition}
\begin{array}{c}\scalebox{1}{\Qcircuit @C=.4em @R=.1em @!R {
&\qw&\qw&\qw&\ctrl{2}&\qw&\qw&\qw&\ctrl{2}&\qw&\ctrl{1}&\gate{T}&\ctrl{1}&\qw\\
&\qw&\ctrl{1}&\qw&\qw&\qw&\ctrl{1}&\qw&\qw&\gate{T}&\targ&\gate{T^\dagger}&\targ&\qw\\
&\gate{H}&\targ&\gate{T^\dagger}&\targ&\gate{T}&\targ&\gate{T^\dagger}&\targ&\gate{T}&\gate{H}&\qw&\qw&\qw
}}\end{array}.
\end{equation}
The $T$-count of Circ.~(\ref{eq:toffoli_standard_decomposition}) is 7. Although Circ.~(\ref{eq:toffoli_standard_decomposition}) appears to have $T$-depth 5, it can be easily transformed into the following quantum circuit, which has $T$-depth 4:
\begin{equation}\label{eq:toffoli_modified_standard_decomposition}
\begin{array}{c}\scalebox{1}{\Qcircuit @C=.4em @R=.1em @!R {
&\qw&\qw&\qw&\ctrl{2}&\qw&\qw&\qw&\ctrl{2}&\ctrl{1}&\gate{T}&\ctrl{1}&\qw\\
&\qw&\ctrl{1}&\qw&\qw&\qw&\ctrl{1}&\gate{T}&\qw&\targ&\gate{T^\dagger}&\targ&\qw\\
&\gate{H}&\targ&\gate{T^\dagger}&\targ&\gate{T}&\targ&\gate{T^\dagger}&\targ&\qw&\gate{T}&\gate{H}&\qw
}}\end{array}.
\end{equation}

Amy \textit{et al.}~\cite{amy2013meet} found the following Toffoli gate decomposition, which keeps $T$-count 7 and improves the $T$-depth to 3:
\begin{equation}\label{eq:toffoli_amy_decomposition}
\begin{array}{c}\scalebox{1}{\Qcircuit @C=.4em @R=.2em @!R {
&\qw&\gate{T}&\targ&\qw&\ctrl{2}&\qw&\ctrl{1}&\gate{T^\dagger}&\qw&\ctrl{2}&\targ&\qw\\
&\qw&\gate{T}&\ctrl{-1}&\targ&\qw&\gate{T^\dagger}&\targ&\gate{T^\dagger}&\targ&\qw&\ctrl{-1}&\qw\\
&\gate{H}&\gate{T}&\qw&\ctrl{-1}&\targ&\qw&\qw&\gate{T}&\ctrl{-1}&\targ&\gate{H}&\qw
}}\end{array}.
\end{equation}
Under the condition of no auxiliary qubit, no decomposition of the Toffoli gate with $T$-depth lower than Circ.~(\ref{eq:toffoli_amy_decomposition}) has been discovered yet.

Selinger~\cite{selinger2013quantum} discovered a CCZ gate decomposition that improves the $T$-depth to 1 while keeping $T$-count 7 by using auxiliary qubits. This CCZ gate decomposition is as follows:
\begin{equation}\label{eq:toffoli_selinger_decomposition}
\begin{array}{c}\scalebox{1}{\Qcircuit @C=.2em @R=.2em @!R {
&&\qw&\qw&\ctrl{3}&\qw&\qw&\qw&\ctrl{4}&\qw&\qw&\gate{T}&\qw&\qw&\ctrl{4}&\qw&\qw&\qw&\ctrl{3}&\qw&\qw&\qw\\
&&\qw&\ctrl{4}&\qw&\ctrl{3}&\qw&\qw&\qw&\qw&\qw&\gate{T}&\qw&\qw&\qw&\qw&\qw&\ctrl{3}&\qw&\ctrl{4}&\qw&\qw\\
&&\qw&\qw&\qw&\qw&\ctrl{3}&\qw&\qw&\ctrl{4}&\qw&\gate{T}&\qw&\ctrl{4}&\qw&\qw&\ctrl{3}&\qw&\qw&\qw&\qw&\qw\\
&\push{\ket{0}}&&\qw&\targ&\qw&\qw&\ctrl{3}&\qw&\qw&\targ&\gate{T}&\targ&\qw&\qw&\ctrl{3}&\qw&\qw&\targ&\qw&\qw&\push{\ket{0}}\\
&\push{\ket{0}}&&\qw&\qw&\targ&\qw&\qw&\targ&\qw&\qw&\gate{T^\dagger}&\qw&\qw&\targ&\qw&\qw&\targ&\qw&\qw&\qw&\push{\ket{0}}\\
&\push{\ket{0}}&&\targ&\qw&\qw&\targ&\qw&\qw&\qw&\ctrl{-2}&\gate{T^\dagger}&\ctrl{-2}&\qw&\qw&\qw&\targ&\qw&\qw&\targ&\qw&\push{\ket{0}}\\
&\push{\ket{0}}&&\qw&\qw&\qw&\qw&\targ&\qw&\targ&\qw&\gate{T^\dagger}&\qw&\targ&\qw&\targ&\qw&\qw&\qw&\qw&\qw&\push{\ket{0}}
}}\end{array}.
\end{equation}
Since the CCZ gate is not a Clifford gate, any Clifford+$T$ gate decomposition of CCZ must include at least one $T$ gate. Therefore, this $T$-depth is the minimum achievable for any Toffoli gate decomposition.

So far, we have discussed the Toffoli gate, i.e., a double-controlled Pauli gate. In contrast, it has been proven that a Clifford+$T$ gate decomposition of a multi-controlled Pauli gate, with more than two control qubits, and without auxiliary qubits, does not exist~\cite{nie2024quantum}.

Selinger~\cite{selinger2013quantum} also demonstrated that for any unitary gate $U$, a double-controlled $U$ gate can be decomposed into a single-controlled $U$ gate and two ``double-controlled $\pm iX$ gates'' as
\begin{equation}\label{eq:c2u_selinger}
\begin{array}{c}\scalebox{1}{\Qcircuit @C=.5em @R=0em @!R {
&\qw&\ctrl{1}&\qw&\qw&&&\qw&\ctrl{1}&\qw&\ctrl{1}&\qw&\qw\\
&\qw&\ctrl{2}&\qw&\qw&\push{\hspace{.34em}=\hspace{-.17em}}&&\qw&\ctrl{1}&\qw&\ctrl{1}&\qw&\qw\\
&&&&&&\push{\ket{0}}&&\gate{-iX}&\ctrl{1}&\gate{iX}&\qw&\push{\ket{0}}\\
&{/}\qw&\gate{U}&\qw&\qw&&&{/\hspace{1.2em}}\qw&\qw&\gate{U}&\qw&\qw&\qw
}}\end{array}.
\end{equation}
Here, the double-controlled $-iX$ gate is defined in their paper as follows:
\begin{equation}\label{eq:ccix_selinger_minus}
\begin{array}{c}\scalebox{1}{\Qcircuit @C=.35em @R=.2em @!R {
&\ctrl{1}&\qw&&&\qw&\qw&\ctrl{3}&\targ&\qw&\gate{T^\dagger}&\qw&\targ&\ctrl{3}&\qw&\qw&\qw\\
&\ctrl{1}&\qw&\push{\coloneqq}&&\qw&\targ&\qw&\qw&\ctrl{2}&\gate{T^\dagger}&\ctrl{2}&\qw&\qw&\targ&\qw&\qw\\
&\gate{-iX}&\qw&&&\gate{H}&\ctrl{-1}&\qw&\ctrl{-2}&\qw&\gate{T}&\qw&\ctrl{-2}&\qw&\ctrl{-1}&\gate{H}&\qw\\
&&&&&\push{\ket{0}\hspace{-1em}}&&\targ&\qw&\targ&\gate{T}&\targ&\qw&\targ&\qw&\push{\hspace{-1em}\ket{0}}&
}}\end{array}.
\end{equation}
While the decomposition of the double-controlled $iX$ gate is not defined in their paper, it likely represents the inverse operation of \cref{eq:ccix_selinger_minus} shown next:
\begin{equation}\label{eq:ccix_selinger_plus}
\begin{array}{c}\scalebox{1}{\Qcircuit @C=.35em @R=.2em @!R {
&\ctrl{1}&\qw&&&\qw&\qw&\ctrl{3}&\targ&\qw&\gate{T}&\qw&\targ&\ctrl{3}&\qw&\qw&\qw\\
&\ctrl{1}&\qw&\push{\coloneqq}&&\qw&\targ&\qw&\qw&\ctrl{2}&\gate{T}&\ctrl{2}&\qw&\qw&\targ&\qw&\qw\\
&\gate{iX}&\qw&&&\gate{H}&\ctrl{-1}&\qw&\ctrl{-2}&\qw&\gate{T^\dagger}&\qw&\ctrl{-2}&\qw&\ctrl{-1}&\gate{H}&\qw\\
&&&&&\push{\ket{0}\hspace{-1em}}&&\targ&\qw&\targ&\gate{T^\dagger}&\targ&\qw&\targ&\qw&\push{\hspace{-1em}\ket{0}}&
}}\end{array}.
\end{equation}
By recursively using \cref{eq:c2u_selinger}, gates with any number of control qubits can be constructed. Furthermore, as illustrated below, they have designed decompositions that execute the double-controlled $\pm iX$ gates in parallel, achieving logarithmic order $T$-depth relative to the number of control qubits:
\begin{equation}\label{eq:c4u_selinger}
\begin{array}{c}\scalebox{1}{\Qcircuit @C=.4em @R=0em @!R {
&\qw&\ctrl{1}&\qw&\qw&&&\qw&\ctrl{1}&\qw&\qw&\qw&\ctrl{1}&\qw&\qw\\
&\qw&\ctrl{2}&\qw&\qw&&&\qw&\ctrl{1}&\qw&\qw&\qw&\ctrl{1}&\qw&\qw\\
&&&&&&\push{\ket{0}}&&\gate{-iX}&\ctrl{3}&\qw&\ctrl{3}&\gate{iX}&\qw&\push{\ket{0}}\\
&\qw&\ctrl{1}&\qw&\qw&&&\qw&\ctrl{1}&\qw&\qw&\qw&\ctrl{1}&\qw&\qw\\
&\qw&\ctrl{3}&\qw&\qw&\push{\raisebox{1em}{\hspace{.34em}=\hspace{-.17em}}}&&\qw&\ctrl{1}&\qw&\qw&\qw&\ctrl{1}&\qw&\qw\\
&&&&&&\push{\ket{0}}&&\gate{-iX}&\ctrl{1}&\qw&\ctrl{1}&\gate{iX}&\qw&\push{\ket{0}}\\
&&&&&&\push{\ket{0}}&&\qw&\gate{-iX}&\ctrl{1}&\gate{iX}&\qw&\qw&\push{\ket{0}}\\
&{/}\qw&\gate{U}&\qw&\qw&&&{/\hspace{1.2em}}\qw&\qw&\qw&\gate{U}&\qw&\qw&\qw&\qw
}}\end{array}.
\end{equation}
They demonstrated that an $n$-controlled NOT gate ($n \geq 3$) can be constructed with a Clifford+$T$ gate decomposition, achieving $T$-count $8n-9$ and $T$-depth $2\lfloor\log_2 (n-2)\rfloor + 3$. For example, this yields an implementation of a triple-controlled NOT gate with $T$-count 15 and $T$-depth 3, and a quintuple-controlled NOT gate with $T$-count 31 and $T$-depth 5.

Jones~\cite{jones2013low} improved the $T$-count for $n$-controlled NOT gates ($n\geq 1$) to about half of Selinger's~\cite{selinger2013quantum} by employing measurement-based feedback control, which uses auxiliary qubits and measures their state to control subsequent quantum circuit operations. This decomposition utilizes \cref{eq:ccix_selinger_minus} as follows:
\begin{equation}\label{eq:jones_add_control}
\begin{array}{c}\scalebox{1}{\Qcircuit @C=.5em @R=0em @!R {
&\qw&\ctrl{1}&\qw&\qw&&&\qw&\ctrl{1}&\qw&\qw&\qw&\qw&\ctrl{1}&\qw&\qw\\
&\qw&\ctrl{2}&\qw&\qw&\push{\hspace{.34em}=\hspace{-.17em}}&&\qw&\ctrl{1}&\qw&\qw&\qw&\qw&\control\qw&\qw&\qw\\
&&&&&&\push{\ket{0}}&&\gate{-iX}&\gate{S}&\ctrl{1}&\gate{H}&\meter&\cctrl{-1}&&\\
&{/}\qw&\gate{U}&\qw&\qw&&&{/\hspace{1.2em}}\qw&\qw&\qw&\gate{U}&\qw&\qw&\qw&\qw&\qw
}}\end{array}.
\end{equation}
\Cref{eq:jones_add_control} shows that adding an additional control qubit to a single-controlled $U$ gate increases the $T$-count by 4. By substituting a CNOT gate for the single-controlled $U$ gate in \cref{eq:jones_add_control}, a decomposition of the Toffoli gate with $T$-count 4 can be constructed. Moreover, by repeatedly adding control qubits to a single-controlled Pauli gate using \cref{eq:jones_add_control}, a decomposition of an $n$-controlled Pauli gate ($n \geq 1$) with $T$-count $4n - 4$ can be constructed.

Gidney~\cite{gidney2018halving} used a different approach from Jones~\cite{jones2013low} to find a method for decomposing an $n$-controlled NOT gate ($n \geq 1$) with $T$-count $4n - 4$. This method is explained below. First, they proposed a quantum gate called a ``temporary logical-AND gate.'' The Clifford+$T$ gate decomposition of the temporary logical-AND gate is as follows:
\begin{equation}\label{eq:logical_and}
\begin{array}{c}\scalebox{1}{\Qcircuit @C=.5em @R=.2em @!R {
&\ctrl{1}&\qw&&&\qw&\qw&\qw&\qw&\ctrl{2}&\targ&\gate{T^\dagger}&\targ&\qw&\qw&\qw\\
&\ctrl{1}&\qw&\push{\hspace{.34em}\coloneqq\hspace{-.17em}}&&\qw&\qw&\qw&\ctrl{1}&\qw&\targ&\gate{T^\dagger}&\targ&\qw&\qw&\qw\\
&&\qw&&\push{\ket{0}}&&\gate{H}&\gate{T}&\targ&\targ&\ctrl{-2}&\gate{T}&\ctrl{-2}&\gate{H}&\gate{S}&\qw
}}\end{array}.
\end{equation}
This is called a ``logical-AND'' because when both the top and middle qubits are $\ket{1}$, the bottom qubit becomes $\ket{1}$, and when either the top or middle qubit is $\ket{0}$, the bottom qubit becomes $\ket{0}$. There is a corresponding uncomputation gate for the temporary logical-AND gate, which can be represented by a Clifford+$T$ gate decomposition as follows:
\begin{equation}\label{eq:uncomputation_logical_and}
\begin{array}{c}\scalebox{1}{\Qcircuit @C=.5em @R=0em @!R {
&\ctrl{1}&\qw&&&\qw&\qw&\ctrl{1}&\qw\\
&\ctrl{1}&\qw&\push{\coloneqq}&&\qw&\qw&\control\qw&\qw\\
&\qw&&&&\gate{H}&\meter&\cctrl{-1}&
}}\end{array},
\end{equation}
where dual-line ``$\begin{array}{c}\Qcircuit @C=3em @R=1em @!R {&\cw}\end{array}$'' represents classical bit,
\begin{equation}
\begin{array}{c}\Qcircuit @C=.5em @R=0em @!R {
&{/}\qw&\gate{U}&\qw\\
&\cw&\cctrl{-1}&\cw
}\end{array}
\coloneqq
\begin{cases}
\begin{array}{c}\Qcircuit @C=.7em @R=0em @!R {
&{/}\qw&\qw&\qw&\qw
}\end{array} & \text{for $\forall U$ if classical bit is 0,}\\
\begin{array}{c}\Qcircuit @C=.5em @R=0em @!R {
&{/}\qw&\gate{U}&\qw
}\end{array} & \text{for $\forall U$ if classical bit is 1.}
\end{cases}
\end{equation}
Using the temporary logical-AND gate and its uncomputation gate, the following equation holds for any unitary gate $U$:
\begin{equation}\label{eq:logical_and_feature}
\begin{array}{c}\scalebox{1}{\Qcircuit @C=.5em @R=0em @!R {
&\qw&\ctrl{1}&\qw&\qw&&&\qw&\ctrl{1}&\qw&\ctrl{1}&\qw&\qw\\
&\qw&\ctrl{2}&\qw&\qw&\push{=}&&\qw&\ctrl{1}&\qw&\ctrl{1}&\qw&\qw\\
&&&&&&&&&\ctrl{1}&\qw&&\\
&{/}\qw&\gate{U}&\qw&\qw&&&{/}\qw&\qw&\gate{U}&\qw&\qw&\qw
}}\end{array}.
\end{equation}
\Cref{eq:logical_and_feature} shows that by using the temporary logical-AND gate and its uncomputation gate, a double-controlled $U$ gate can be represented by a single-controlled $U$ gate. They demonstrated that by repeatedly adding control qubits to the CZ gate $n-1$ times, as described in \cref{eq:logical_and_feature}, a Clifford+$T$ gate decomposition of a C$^n$Z gate with $T$-count $4n-4$ can be constructed.

Subsequently, Gidney \textit{et al.}~\cite{gidney2021cccz} discovered a further improved $T$-count for the Clifford+$T$ gate decomposition of the C$^n$Z gate. First, they found that a CCCZ gate can be constructed with $T$-count just 6, as follows:
\begin{equation}\label{eq:gidney_cccz}
\begin{array}{c}\scalebox{.8}{\Qcircuit @C=.2em @R=-.3em @!R {
&&\qw&\qw&\qw&\qw&\qw&\ctrl{4}&\qw&\qw&\qw&\qw&\qw&\qw&\qw&\qw&\qw&\qw&\qw&\qw&\control\qw&\qw&\qw\\
&&\qw&\qw&\qw&\ctrl{3}&\qw&\qw&\qw&\ctrl{3}&\qw&\qw&\qw&\qw&\qw&\qw&\qw&\qw&\qw&\qw&\ctrl{-1}&\qw&\qw\\
&&\qw&\qw&\qw&\qw&\qw&\qw&\qw&\qw&\ctrl{2}&\qw&\qw&\qw&\ctrl{2}&\qw&\qw&\qw&\qw&\ctrl{1}&\qw&\qw&\qw\\
&&\qw&\qw&\qw&\qw&\qw&\qw&\qw&\qw&\qw&\qw&\ctrl{1}&\qw&\qw&\qw&\ctrl{1}&\qw&\qw&\control\qw&\qw&\qw&\qw\\
&\push{\ket{0}}&&\gate{H}&\gate{T}&\targ&\gate{T^\dagger}&\targ&\gate{T}&\targ&\targ&\gate{T^\dagger}&\targ&\gate{T}&\targ&\gate{T^\dagger}&\targ&\gate{\sqrt{X}^\dagger}&\meter&\cctrlo{-1}&\cctrl{-3}&&
}}\end{array}\!.\!\!\!\!
\end{equation}
By applying the control qubit addition method, \cref{eq:logical_and_feature}, to the CCCZ gate decomposition~(\ref{eq:gidney_cccz}) $n-3$ times, they showed that a Clifford+$T$ gate decomposition of a C$^n$Z gate ($n\geq 3$) with $T$-count $4n-6$ can be constructed. This is currently the known Clifford+$T$ gate decomposition of a C$^n$Z gate with the smallest $T$-count.

Paler \textit{et al.}~\cite{paler2022realistic} noted that the temporary logical-AND gate shown in \cref{eq:logical_and} is one of the relative phase Toffoli gates. The relative phase Toffoli gate performs a Toffoli operation but differs by inducing a relative phase shift~\cite{Barenco1995-ro,nielsen2010quantum}. The double-controlled $\pm iX$ gates shown in \cref{eq:ccix_selinger_minus,eq:ccix_selinger_plus} are also examples of the relative phase Toffoli gates. They used several different relative phase Toffoli gates, one of which is shown below:
\begin{equation}\label{eq:paler_rtof}
\begin{array}{c}\scalebox{1}{\Qcircuit @C=.5em @R=0em @!R {
&\qw&\qw&\qw&\qw&\ctrl{2}&\qw&\qw&\qw&\qw&\qw\\
&\qw&\qw&\ctrl{1}&\qw&\qw&\qw&\ctrl{1}&\qw&\qw&\qw\\
&\gate{H}&\gate{T}&\targ&\gate{T^\dagger}&\targ&\gate{T}&\targ&\gate{T^\dagger}&\gate{H}&\qw
}}\end{array}.
\end{equation}
They found a Toffoli gate decomposition with CNOT-count 4 or 5 while keeping $T$-count 4, by using Circ.~(\ref{eq:paler_rtof}) instead of the temporary logical-AND gate shown in \cref{eq:logical_and} and changing some of the gates accordingly:
\begin{equation}\label{eq:paler_toffoli}
\begin{array}{c}\scalebox{.8}{\Qcircuit @C=.3em @R=0em @!R {
&&\qw&\qw&\qw&\qw&\qw&\ctrl{2}&\qw&\qw&\qw&\qw&\qw&\qw&\qw&\qw&\ctrl{1}&\qw&\qw\\
&&\qw&\qw&\qw&\ctrl{1}&\qw&\qw&\qw&\ctrl{1}&\qw&\qw&\qw&\qw&\qw&\qw&\control\qw&\qw&\qw\\
&\push{\ket{0}}&&\gate{H}&\gate{T}&\targ&\gate{T^\dagger}&\targ&\gate{T}&\targ&\gate{T^\dagger}&\gate{H}&\gate{S^\dagger}&\ctrl{1}&\gate{H}&\meter&\cctrl{-1}&&\\
&&\qw&\qw&\qw&\qw&\qw&\qw&\qw&\qw&\qw&\qw&\qw&\targ&\qw&\qw&\qw&\qw&\qw
}}\end{array}.
\end{equation}

\section{Fewer \texorpdfstring{$T$}{T}-depth decomposition without measurement-based feedback control}\label{sec:improve_selinger}

Before moving on to the main topic of this paper, we show that the $T$-depth of the Clifford+$T$ gate decomposition of multi-controlled Pauli gates, as described by Selinger~\cite{selinger2013quantum}, can be improved a little more. For example, Selinger~\cite{selinger2013quantum} claims that a quintuple-controlled NOT gate can be constructed with $T$-count 31 and $T$-depth 5, but we show that this $T$-depth can be reduced to 3. First, using Circ.~(\ref{eq:toffoli_selinger_decomposition}) and \cref{eq:c2u_selinger,eq:ccix_selinger_minus,eq:ccix_selinger_plus}, C$^5$Z can be decomposed as follows:
\begin{equation}\label{eq:c5z_without_measure}
\begin{array}{c}\scalebox{1}{\Qcircuit @C=.5em @R=-.2em @!R {
&&\qw&\ctrl{1}&\qw&\ctrl{1}&\qw&\qw\\
&&\qw&\ctrl{1}&\qw&\ctrl{1}&\qw&\qw\\
&\push{\ket{0}}&&\gate{-iX}&\ctrl{3}&\gate{iX}&\qw&\push{\ket{0}}\\
&&\qw&\ctrl{1}&\qw&\ctrl{1}&\qw&\qw\\
&&\qw&\ctrl{1}&\qw&\ctrl{1}&\qw&\qw\\
&\push{\ket{0}}&&\gate{-iX}&\ctrl{3}&\gate{iX}&\qw&\push{\ket{0}}\\
&&\qw&\ctrl{1}&\qw&\ctrl{1}&\qw&\qw\\
&&\qw&\ctrl{1}&\qw&\ctrl{1}&\qw&\qw\\
&\push{\ket{0}}&&\gate{-iX}&\control\qw&\gate{iX}&\qw&\push{\ket{0}}
}}\end{array}.
\end{equation}
Since Circ.~(\ref{eq:toffoli_selinger_decomposition}) and \cref{eq:ccix_selinger_minus,eq:ccix_selinger_plus} each have $T$-depth 1, a C$^5$Z gate decomposition can be constructed with $T$-depth 3. From Circ.~(\ref{eq:c5z_without_measure}) and \cref{eq:cnx_cny_cnz}, a quintuple-controlled Pauli gate decomposition can also be constructed with $T$-depth 3. Similarly, by applying \cref{eq:ccix_selinger_minus,eq:ccix_selinger_plus} to $k$ qubits $(1\leq k\leq n)$ in the C$^{m-1}$Z gate, it is possible to construct C$^{m+k-1}$Z gate by only increasing the $T$-depth by 2. Consequently, the $T$-depth of $n$-controlled Pauli gates ($n\geq 3$) can be reduced to $2\lfloor\log_2 (\frac{n}{3})\rfloor + 3$ while keeping $T$-count $8n-9$.

\section{Fewer CNOT-count / \texorpdfstring{$T$}{T}-depth decompositions}\label{sec:main_decompositions}

In this section, we describe our proposed method for constructing Clifford+$T$ gate decompositions of multi-controlled Pauli gates. The following equation is the key equation in this paper for constructing Clifford+$T$ gate decompositions of multi-controlled Pauli gates:
\begin{equation}\label{eq:cnz_base}
\begin{array}{c}\scalebox{1}{\Qcircuit @C=.6em @R=0em @!R {
&{/^m}\qw&\qw&\qw&\gate{U}&\qw&\qw&\qw&\qw&\qw&&&{/^m}\qw&\qw&\gate{U}&\qw&\qw\\
&{/^n}\qw&\qw&\ctrl{1}&\qw&\qw&\qw&\control\qw&\qw&\qw&\push{\raisebox{1em}{=}}&&{/^n}\qw&\qw&\ctrl{-1}&\qw&\qw\\
&&\lstick{\ket{0}}&\targ&\ctrl{-2}&\gate{H}&\meter&\cctrl{-1}&&&&&&&&&
}}\end{array},
\end{equation}
where $m,n\in\mathbb{Z}_{>0}$, $U$ denotes any $m$-qubit unitary gate, and
\begin{equation}
\begin{array}{c}\Qcircuit @C=1em @R=1em @!R {
&{/^n}\qw&\qw&\control\qw&\qw&\qw
}\end{array}
\coloneqq
\begin{cases}
\begin{array}{c}\Qcircuit @C=1.5em @R=1em @!R {
&\gate{Z}&\qw
}\end{array} & \text{if $n=1$,}\bigskip\\
\begin{array}{c}\Qcircuit @C=1em @R=1em @!R {
&\qw&\ctrl{1}&\qw&\qw\\
&\qw&\ctrl{2}&\qw&\qw\\
&\raisebox{.3em}{\ensuremath{\vdots}}&&\raisebox{.3em}{\ensuremath{\vdots}}&\\
&\qw&\control\qw&\qw&\qw
\gategroup{1}{5}{4}{5}{.5em}{)}
}\end{array}\raisebox{-.3em}{\text{$n$ qubits}} & \text{if $n \geq 2$.}
\end{cases}
\end{equation}
The proof of \cref{eq:cnz_base} is in the Appendix. In \cref{sec:few_cnot_count,sec:few_t_depth}, we use \cref{eq:cnz_base} to construct the following three:
\begin{itemize}
\item CCZ gate decomposition,
\item CCCZ gate decomposition,
\item Control qubit addition method.
\end{itemize}

Before starting the detailed discussion, we introduces some equations that help you when reading this section.

The equation below demonstrates that when the input is $\ket{0}$, any subsequent phase gate $P(\theta)$ (e.g., $Z$ gate, $S$ gate, $T$ gate, $S^\dagger$ gate, and $T^\dagger$ gate) leaves the result unchanged:
\begin{equation}\label{eq:ket0_phase}
\begin{array}{c}\Qcircuit @C=1em @R=0em @!R {
\lstick{\ket{0}}&\gate{P(\theta)}&\qw&\push{=}&&&\lstick{\ket{0}}&\qw&\qw&\qw&\qw
}\end{array}\quad\text{for $\forall\theta$.}
\end{equation}

Similarly, when the input is $\ket{0}$, the subsequent controlled $U$ gate, as shown below, does not change the result either for any unitary gate $U$:
\begin{equation}\label{eq:ket0_control}
\begin{array}{c}\Qcircuit @C=1em @R=0em @!R {
&{/}\qw&\gate{U}&\qw&&&&&{/}\qw&\qw&\qw&\qw\\
\lstick{\ket{0}}&\qw&\ctrl{-1}&\qw&\push{\raisebox{1em}{=}}&&&\lstick{\ket{0}}&\qw&\qw&\qw&\qw
}\end{array}\quad\text{for $\forall U$.}
\end{equation}

The following equation demonstrates that a phase gate $P(\theta)$ applied just before a measurement does not affect the result:
\begin{equation}\label{eq:phase_measure}
\begin{array}{c}\Qcircuit @C=1em @R=0em @!R {
&\gate{P(\theta)}&\meter&\cw&\push{=}&&\qw&\meter&\cw&\cw
}\end{array}\quad\text{for $\forall\theta$.}
\end{equation}

The following equations hold with respect to the controlled $U$ gate and the subsequent measurement for any unitary gate $U$:
\begin{gather}
\begin{array}{c}\Qcircuit @C=.7em @R=0em @!R {
&{/}\qw&\gate{U}&\qw&\qw&&&{/}\qw&\qw&\gate{U}&\qw\\
&\qw&\ctrl{-1}&\meter&\cw&\push{\raisebox{1em}{=}}&&\qw&\meter&\cctrl{-1}&\cw
}\end{array}\quad\text{for $\forall U$,}\label{eq:control_measure}\\
\begin{array}{c}\Qcircuit @C=.7em @R=0em @!R {
&{/}\qw&\gate{U}&\qw&\qw&&&{/}\qw&\qw&\gate{U}&\qw\\
&\qw&\ctrlo{-1}&\meter&\cw&\push{\raisebox{1em}{=}}&&\qw&\meter&\cctrlo{-1}&\cw
}\end{array}\quad\text{for $\forall U$,}\label{eq:controlo_measure}
\end{gather}
where
\begin{gather}
\begin{array}{c}\Qcircuit @C=.5em @R=0em @!R {
&{/}\qw&\gate{U}&\qw&&&{/}\qw&\qw&\gate{U}&\qw&\qw\\
&\qw&\ctrlo{-1}&\qw&\push{\raisebox{1em}{$\coloneqq$}}&&\qw&\gate{X}&\ctrl{-1}&\gate{X}&\qw
}\end{array}\quad\text{for $\forall U$,}\\
\begin{array}{c}\Qcircuit @C=.5em @R=0em @!R {
&{/}\qw&\gate{U}&\qw\\
&\cw&\cctrlo{-1}&\cw
}\end{array}
\coloneqq
\begin{cases}
\begin{array}{c}\Qcircuit @C=.5em @R=0em @!R {
&{/}\qw&\gate{U}&\qw
}\end{array} & \text{for $\forall U$ if classical bit is 0,}\\
\begin{array}{c}\Qcircuit @C=.7em @R=0em @!R {
&{/}\qw&\qw&\qw&\qw
}\end{array} & \text{for $\forall U$ if classical bit is 1.}
\end{cases}
\end{gather}

\subsection{Fewer CNOT-count decompositions}\label{sec:few_cnot_count}

\begin{figure}[t]
\vspace{3.8ex}
\flushleft{(a)}\vspace{-3.8ex}
\begin{equation*}
\begin{array}{c}\scalebox{.8}{\Qcircuit @C=.3em @R=-.1em @!R {
&&\qw&\qw&\qw&\qw&\qw&\qw&\qw&\qw&\qw&\ctrl{3}&\qw&\qw&\qw&\qw&\qw&\qw\\
&&\qw&\qw&\qw&\qw&\qw&\ctrl{2}&\qw&\qw&\qw&\qw&\qw&\qw&\qw&\ctrl{1}&\qw&\qw\\
&&\qw&\qw&\qw&\ctrl{1}&\qw&\qw&\qw&\ctrl{1}&\qw&\qw&\qw&\qw&\qw&\control\qw&\qw&\qw\\
&\push{\ket{0}}&&\gate{H}&\gate{T}&\targ&\gate{T^\dagger}&\targ&\gate{T}&\targ&\gate{T^\dagger}&\targ&\gate{S}&\gate{H}&\meter&\cctrl{-1}&&
}}\end{array}\smallskip
\end{equation*}
\flushleft{(b)}\vspace{-3.8ex}
\begin{equation*}
\begin{array}{c}\scalebox{.8}{\Qcircuit @C=.3em @R=.2em @!R {
&&\qw&\qw&\qw&\qw&\qw&\qw&\qw&\qw&\ctrl{3}&\qw&\qw&\qw&\qw&\qw&\qw\\
&&\qw&\qw&\qw&\qw&\ctrl{2}&\targ&\gate{T^\dagger}&\targ&\qw&\qw&\qw&\qw&\ctrl{1}&\qw&\qw\\
&&\qw&\qw&\qw&\ctrl{1}&\qw&\targ&\gate{T^\dagger}&\targ&\qw&\qw&\qw&\qw&\control\qw&\qw&\qw\\
&\push{\ket{0}}&&\gate{H}&\gate{T}&\targ&\targ&\ctrl{-2}&\gate{T}&\ctrl{-2}&\targ&\gate{S}&\gate{H}&\meter&\cctrlo{-1}&&
}}\end{array}\smallskip
\end{equation*}
\flushleft{(c)} \vspace{-3.8ex}
\begin{equation*}
\begin{array}{c}\scalebox{.8}{\Qcircuit @C=.3em @R=.2em @!R {
&&\qw&\qw&\qw&\qw&\qw&\qw&\qw&\qw&\qw&\ctrl{3}&\qw&\qw&\qw&\qw&\qw&\qw\\
&&\qw&\qw&\ctrl{3}&\targ&\qw&\gate{T^\dagger}&\qw&\targ&\ctrl{3}&\qw&\qw&\qw&\qw&\ctrl{1}&\qw&\qw\\
&&\qw&\qw&\qw&\targ&\ctrl{2}&\gate{T^\dagger}&\ctrl{2}&\targ&\qw&\qw&\qw&\qw&\qw&\control\qw&\qw&\qw\\
&\push{\ket{0}}&&\gate{H}&\qw&\ctrl{-2}&\qw&\gate{T}&\qw&\ctrl{-2}&\qw&\targ&\gate{S}&\gate{H}&\meter&\cctrlo{-1}&&\\
&&\push{\ket{0}\hspace{-.8em}}&&\targ&\qw&\targ&\gate{T}&\targ&\qw&\targ&\qw&\push{\hspace{-1em}\ket{0}}&&&&&
}}\end{array}
\end{equation*}
\caption{
CCZ gate decompositions with $T$-count 4.
(a) Discovered by Paler \textit{et al.}~\cite{paler2022realistic}. We also derived in \cref{sec:few_cnot_count}.
(b) Discovered by Gidney~\cite{gidney2018halving}. We also derived in \cref{sec:few_t_depth}.
(c) Discovered by Jones~\cite{jones2013low}. We also derived in \cref{sec:few_t_depth}.
}
\label{fig:comp_ccz}
\end{figure}

\begin{figure}[t]
\vspace{3.8ex}
\flushleft{(a)}\vspace{-6ex}
\begin{equation*}
\begin{array}{c}\scalebox{.8}{\Qcircuit @C=.3em @R=-.1em @!R {
&&{/}\qw&\qw&\qw&\qw&\qw&\qw&\qw&\qw&\qw&\qw&\gate{U}&\qw&\qw&\qw&\qw&\qw&\qw&&&{/}\qw&\qw&\gate{U}&\qw&\qw\\
&&\qw&\qw&\qw&\qw&\qw&\ctrl{2}&\qw&\qw&\qw&\qw&\qw&\qw&\qw&\qw&\ctrl{1}&\qw&\qw&\push{=}&&\qw&\qw&\ctrl{-1}&\qw&\qw\\
&&\qw&\qw&\qw&\ctrl{1}&\qw&\qw&\qw&\ctrl{1}&\qw&\qw&\qw&\qw&\qw&\qw&\control\qw&\qw&\qw&&&\qw&\qw&\ctrl{-1}&\qw&\qw\\
&\push{\ket{0}}&&\gate{H}&\gate{T}&\targ&\gate{T^\dagger}&\targ&\gate{T}&\targ&\gate{T^\dagger}&\gate{H}&\ctrl{-3}&\gate{S}&\gate{H}&\meter&\cctrlo{-1}&&&&&&&&&
}}\end{array}\smallskip
\end{equation*}
\flushleft{(b)}\vspace{-6ex}
\begin{equation*}
\begin{array}{c}\scalebox{.8}{\Qcircuit @C=.3em @R=.2em @!R {
&&{/}\qw&\qw&\qw&\qw&\qw&\qw&\qw&\qw&\qw&\gate{U}&\qw&\qw&\qw&\qw&\qw&\qw&&&{/}\qw&\qw&\gate{U}&\qw&\qw\\
&&\qw&\qw&\qw&\qw&\ctrl{2}&\targ&\gate{T^\dagger}&\targ&\qw&\qw&\qw&\qw&\qw&\ctrl{1}&\qw&\qw&\push{=}&&\qw&\qw&\ctrl{-1}&\qw&\qw\\
&&\qw&\qw&\qw&\ctrl{1}&\qw&\targ&\gate{T^\dagger}&\targ&\qw&\qw&\qw&\qw&\qw&\control\qw&\qw&\qw&&&\qw&\qw&\ctrl{-1}&\qw&\qw\\
&\push{\ket{0}}&&\gate{H}&\gate{T}&\targ&\targ&\ctrl{-2}&\gate{T}&\ctrl{-2}&\gate{H}&\ctrl{-3}&\gate{S}&\gate{H}&\meter&\cctrl{-1}&&&&&&&&&
}}\end{array}\smallskip
\end{equation*}
\flushleft{(c)} \vspace{-6ex}
\begin{equation*}
\begin{array}{c}\scalebox{.8}{\Qcircuit @C=.3em @R=.2em @!R {
&&{/}\qw&\qw&\qw&\qw&\qw&\qw&\qw&\qw&\qw&\qw&\gate{U}&\qw&\qw&\qw&\qw&\qw&\qw&&&{/}\qw&\qw&\gate{U}&\qw&\qw\\
&&\qw&\qw&\ctrl{3}&\targ&\qw&\gate{T^\dagger}&\qw&\targ&\ctrl{3}&\qw&\qw&\qw&\qw&\qw&\ctrl{1}&\qw&\qw&\push{=}&&\qw&\qw&\ctrl{-1}&\qw&\qw\\
&&\qw&\qw&\qw&\targ&\ctrl{2}&\gate{T^\dagger}&\ctrl{2}&\targ&\qw&\qw&\qw&\qw&\qw&\qw&\control\qw&\qw&\qw&&&\qw&\qw&\ctrl{-1}&\qw&\qw\\
&\push{\ket{0}}&&\gate{H}&\qw&\ctrl{-2}&\qw&\gate{T}&\qw&\ctrl{-2}&\qw&\gate{H}&\ctrl{-3}&\gate{S}&\gate{H}&\meter&\cctrl{-1}&&&&&&&&&\\
&&\push{\ket{0}\hspace{-.8em}}&&\targ&\qw&\targ&\gate{T}&\targ&\qw&\targ&\qw&\push{\hspace{-1.8em}\ket{0}}&&&&&&&&&&&&&
}}\end{array}
\end{equation*}
\caption{
Control qubit addition methods with $T$-count +4. $U$ denotes any unitary gate.
(a) Generalization of the Toffoli gate decomposition by Paler \textit{et al.}~\cite{paler2022realistic}. We derived in \cref{sec:few_cnot_count}.
(b) Discovered by Gidney~\cite{gidney2018halving}. We also derived in \cref{sec:few_t_depth}.
(c) Discovered by Jones~\cite{jones2013low}. We also derived in \cref{sec:few_t_depth}.
}
\label{fig:comp_cadd}
\end{figure}

This subsection explains how to construct multi-controlled Pauli gate decompositions that keep the currently known minimum $T$-count while achieving smaller CNOT-count. There are two points that are particularly worth noting. Firstly, the CCZ gate decomposition, the CCCZ gate decomposition, and the control qubit addition method can all be constructed using almost the same procedure. Secondly, this has improved the CNOT-count of the CCCZ gate decompositions while keeping the $T$-count at the currently known minimum of 6.

We first noticed that Circ.~(\ref{eq:paler_rtof}) can be rewritten as follows:
\begin{equation}\label{eq:rtof_1a}
\begin{array}{c}\scalebox{.8}{\Qcircuit @C=.3em @R=0em @!R {
&\qw&\qw&\qw&\qw&\ctrl{2}&\qw&\qw&\qw&\qw&\qw&&&\ctrl{1}&\qw&\ctrl{1}&\qw&\qw&&&\qw&\ctrl{1}&\qw&\ctrl{1}&\qw\\
&\qw&\qw&\ctrl{1}&\qw&\qw&\qw&\ctrl{1}&\qw&\qw&\qw&\push{=}&&\ctrlo{1}&\qw&\ctrl{1}&\qw&\qw&\push{=}&&\qw&\ctrl{1}&\qw&\ctrlo{1}&\qw\\
&\gate{H}&\gate{T}&\targ&\gate{T^\dagger}&\targ&\gate{T}&\targ&\gate{T^\dagger}&\gate{H}&\qw&&&\control\qw&\gate{S^\dagger}&\targ&\gate{S}&\qw&&&\gate{S^\dagger}&\targ&\gate{S}&\control\qw&\qw
}}\end{array}.
\end{equation}
Likewise, the following equation also holds:
\begin{equation}\label{eq:rtof_1b}
\begin{array}{c}\scalebox{.8}{\Qcircuit @C=.3em @R=0em @!R {
&\qw&\qw&\qw&\qw&\ctrl{2}&\qw&\qw&\qw&\qw&\qw&&&\ctrl{1}&\qw&\ctrl{1}&\qw&\qw&&&\qw&\ctrl{1}&\qw&\ctrl{1}&\qw\\
&\qw&\qw&\ctrl{1}&\qw&\qw&\qw&\ctrl{1}&\qw&\qw&\qw&\push{=}&&\ctrlo{1}&\qw&\ctrl{1}&\qw&\qw&\push{=}&&\qw&\ctrl{1}&\qw&\ctrlo{1}&\qw\\
&\gate{H}&\gate{T^\dagger}&\targ&\gate{T}&\targ&\gate{T^\dagger}&\targ&\gate{T}&\gate{H}&\qw&&&\control\qw&\gate{S}&\targ&\gate{S^\dagger}&\qw&&&\gate{S}&\targ&\gate{S^\dagger}&\control\qw&\qw
}}\end{array}.
\end{equation}
All four decompositions (Figs.~\ref{fig:comp_ccz}a, \ref{fig:comp_cccz}a, \ref{fig:comp_cccz}a', and \ref{fig:comp_cadd}a) are obtained by transforming \cref{eq:cnz_base} and substituting \cref{eq:rtof_1a,eq:rtof_1b}.
\Cref{fig:comp_ccz}a presents a CCZ gate decomposition with $T$-count 4 and CNOT-count either 4 or 5. These $T$-count and CNOT-count are the smallest currently known for such decompositions. However, this double-controlled Pauli gate decomposition was already discovered by Paler \textit{et al.}~\cite{paler2022realistic}, so it lacks novelty. \Cref{fig:comp_cccz}a presents a CCCZ gate decomposition with $T$-count 6 and CNOT-count either 6 or 8. Additionally, \cref{fig:comp_cccz}a' provides a CCCZ gate decomposition with $T$-count 6 and CNOT-count 7. These decompositions successfully improve the CNOT-count over the currently known decomposition~(\ref{eq:gidney_cccz}), which has $T$-count 6 and CNOT-count 8. \Cref{fig:comp_cadd}a shows a control qubit addition method. Using this method, when you add a control qubit to a multi-controlled $Z$ gate, the $T$-count increases by 4 and the CNOT-count increases by 3 or 4. \Cref{fig:comp_cadd}a can be viewed as a generalization of the Toffoli gate decomposition by Paler \textit{et al.}~\cite{paler2022realistic}. Although Fig.~\ref{fig:comp_ccz}a can also be derived from the method of Paler~\cite{paler2022realistic}, Figs.~\ref{fig:comp_cccz}a and \ref{fig:comp_cccz}a' can only be derived using our procedure. Since Figs.~\ref{fig:comp_ccz}a, \ref{fig:comp_cccz}a, \ref{fig:comp_cccz}a', and \ref{fig:comp_cadd}a can all be derived in a similar manner, we provide the derivation of \cref{fig:comp_cccz}a below, and the derivation of \cref{fig:comp_cccz}a' is outlined in \cref{apx:cccz_min_cnot_count_alternative}.

\begin{proof}
By substituting $m=2$ and $n=2$ into \cref{eq:cnz_base}, and setting $U$ as the CZ gate, we find that the following quantum circuit is a decomposition of the CCCZ gate:
\begin{equation}\label{eq:cccz_deriv_1}
\begin{array}{c}\scalebox{1}{\Qcircuit @C=.5em @R=0em @!R {
&&\qw&\qw&\ctrl{1}&\qw&\qw&\qw&\qw&\qw\\
&&\qw&\qw&\ctrl{3}&\qw&\qw&\qw&\qw&\qw\\
&&\qw&\ctrl{1}&\qw&\qw&\qw&\ctrl{1}&\qw&\qw\\
&&\qw&\ctrl{1}&\qw&\qw&\qw&\control\qw&\qw&\qw\\
&\push{\ket{0}}&&\targ&\control\qw&\gate{H}&\meter&\cctrl{-1}&&
}}\end{array}.
\end{equation}
Circuit~(\ref{eq:cccz_deriv_1}) can be transformed using \cref{eq:cnx_cny_cnz} as follows,
\begin{equation}\label{eq:cccz_deriv_2}
\begin{array}{c}\scalebox{1}{\Qcircuit @C=.5em @R=0em @!R {
&&\qw&\qw&\qw&\ctrl{1}&\qw&\qw&\qw&\qw\\
&&\qw&\qw&\qw&\ctrl{3}&\qw&\qw&\qw&\qw\\
&&\qw&\ctrl{1}&\qw&\qw&\qw&\ctrl{1}&\qw&\qw\\
&&\qw&\ctrl{1}&\qw&\qw&\qw&\control\qw&\qw&\qw\\
&\push{\ket{0}}&&\targ&\gate{H}&\targ&\meter&\cctrl{-1}&&
}}\end{array}.
\end{equation}
Circuit~(\ref{eq:cccz_deriv_2}) can be transformed using \cref{eq:ket0_phase,eq:ket0_control,eq:phase_measure} as follows,
\begin{equation}\label{eq:cccz_deriv_3}
\begin{array}{c}\scalebox{1}{\Qcircuit @C=.5em @R=0em @!R {
&&\qw&\qw&\qw&\qw&\qw&\ctrl{1}&\qw&\qw&\qw&\qw&\qw\\
&&\qw&\qw&\qw&\qw&\qw&\ctrl{3}&\qw&\qw&\qw&\qw&\qw\\
&&\qw&\ctrl{1}&\qw&\ctrl{1}&\qw&\qw&\qw&\qw&\ctrl{1}&\qw&\qw\\
&&\qw&\ctrlo{1}&\qw&\ctrl{1}&\qw&\qw&\qw&\qw&\control\qw&\qw&\qw\\
&\push{\ket{0}}&&\control\qw&\gate{S^\dagger}&\targ&\gate{H}&\targ&\gate{S^\dagger}&\meter&\cctrl{-1}&&
}}\end{array}.
\end{equation}
Using the fact that
\begin{equation*}
\begin{array}{c}\Qcircuit @C=1em @R=.5em @!R {
&\ctrl{1}&\ctrl{1}&\qw&&&\qw&\qw\\
&\ctrlo{1}&\ctrlo{1}&\qw&\push{=}&&\qw&\qw\\
&\control\qw&\control\qw&\qw&&&\qw&\qw
}\end{array},
\end{equation*}
Circ.~(\ref{eq:cccz_deriv_3}) can be transformed into
\begin{equation}\label{eq:cccz_deriv_4}
\begin{array}{c}\scalebox{1}{\Qcircuit @C=.5em @R=0em @!R {
&&\qw&\qw&\qw&\qw&\qw&\ctrl{1}&\qw&\ctrl{1}&\ctrl{1}&\qw&\qw&\qw&\qw\\
&&\qw&\qw&\qw&\qw&\qw&\ctrl{3}&\qw&\ctrlo{3}&\ctrlo{3}&\qw&\qw&\qw&\qw\\
&&\qw&\ctrl{1}&\qw&\ctrl{1}&\qw&\qw&\qw&\qw&\qw&\qw&\ctrl{1}&\qw&\qw\\
&&\qw&\ctrlo{1}&\qw&\ctrl{1}&\qw&\qw&\qw&\qw&\qw&\qw&\control\qw&\qw&\qw\\
&\push{\ket{0}}&&\control\qw&\gate{S^\dagger}&\targ&\gate{H}&\targ&\gate{S^\dagger}&\control\qw&\control\qw&\meter&\cctrl{-1}&&
}}\end{array}.
\end{equation}
Circuit~(\ref{eq:cccz_deriv_4}) can be transformed using \cref{eq:control_measure} as follows,
\begin{equation}\label{eq:cccz_deriv_5}
\begin{array}{c}\scalebox{1}{\Qcircuit @C=.5em @R=0em @!R {
&&\qw&\qw&\qw&\qw&\qw&\ctrl{1}&\qw&\ctrl{1}&\qw&\ctrl{1}&\qw&\qw&\qw\\
&&\qw&\qw&\qw&\qw&\qw&\ctrl{3}&\qw&\ctrlo{3}&\qw&\controlo\qw&\qw&\qw&\qw\\
&&\qw&\ctrl{1}&\qw&\ctrl{1}&\qw&\qw&\qw&\qw&\qw&\qw&\ctrl{1}&\qw&\qw\\
&&\qw&\ctrlo{1}&\qw&\ctrl{1}&\qw&\qw&\qw&\qw&\qw&\qw&\control\qw&\qw&\qw\\
&\push{\ket{0}}&&\control\qw&\gate{S^\dagger}&\targ&\gate{H}&\targ&\gate{S^\dagger}&\control\qw&\meter&\cctrl{-3}&\cctrl{-1}&&
}}\end{array}.
\end{equation}
Circuit~(\ref{eq:cccz_deriv_5}) can be transformed using the fact that
\begin{equation*}
\begin{array}{c}\Qcircuit @C=1em @R=.5em @!R {
&\gate{S}&\gate{S^\dagger}&\qw&\push{=}&&\gate{S}&\gate{S^\dagger}&\qw&\push{=}&&\qw&\qw
}\end{array}
\end{equation*}
as follows,
\begin{equation}\label{eq:cccz_deriv_6}
\begin{array}{c}\scalebox{.8}{\Qcircuit @C=.5em @R=0em @!R {
&&\qw&\qw&\qw&\qw&\qw&\qw&\qw&\qw&\qw&\ctrl{1}&\qw&\ctrl{1}&\qw&\ctrl{1}&\qw&\qw&\qw\\
&&\qw&\qw&\qw&\qw&\qw&\qw&\qw&\qw&\qw&\ctrl{3}&\qw&\ctrlo{3}&\qw&\controlo\qw&\qw&\qw&\qw\\
&&\qw&\ctrl{1}&\qw&\ctrl{1}&\qw&\qw&\qw&\qw&\qw&\qw&\qw&\qw&\qw&\qw&\ctrl{1}&\qw&\qw\\
&&\qw&\ctrlo{1}&\qw&\ctrl{1}&\qw&\qw&\qw&\qw&\qw&\qw&\qw&\qw&\qw&\qw&\control\qw&\qw&\qw\\
&\push{\ket{0}}&&\control\qw&\gate{S^\dagger}&\targ&\gate{S}&\gate{S^\dagger}&\gate{H}&\gate{S^\dagger}&\gate{S}&\targ&\gate{S^\dagger}&\control\qw&\meter&\cctrl{-3}&\cctrl{-1}&&
\relax\gategroup{3}{4}{5}{7}{.5em}{--}
\relax\gategroup{1}{11}{5}{14}{.5em}{--}
}}\end{array}.
\end{equation}
Replacing the portion of Circ.~(\ref{eq:cccz_deriv_6}) enclosed by the dashed line with \cref{eq:rtof_1a,eq:rtof_1b}, we get
\begin{equation}\label{eq:cccz_deriv_7}
\begin{array}{c}\scalebox{.5}{\Qcircuit @C=.3em @R=0em @!R {
&&\qw&\qw&\qw&\qw&\qw&\qw&\qw&\qw&\qw&\qw&\qw&\qw&\qw&\qw&\qw&\qw&\qw&\ctrl{4}&\qw&\qw&\qw&\qw&\qw&\ctrl{1}&\qw&\qw&\qw\\
&&\qw&\qw&\qw&\qw&\qw&\qw&\qw&\qw&\qw&\qw&\qw&\qw&\qw&\qw&\qw&\ctrl{3}&\qw&\qw&\qw&\ctrl{3}&\qw&\qw&\qw&\controlo\qw&\qw&\qw&\qw\\
&&\qw&\qw&\qw&\qw&\qw&\ctrl{2}&\qw&\qw&\qw&\qw&\qw&\qw&\qw&\qw&\qw&\qw&\qw&\qw&\qw&\qw&\qw&\qw&\qw&\qw&\ctrl{1}&\qw&\qw\\
&&\qw&\qw&\qw&\ctrl{1}&\qw&\qw&\qw&\ctrl{1}&\qw&\qw&\qw&\qw&\qw&\qw&\qw&\qw&\qw&\qw&\qw&\qw&\qw&\qw&\qw&\qw&\control\qw&\qw&\qw\\
&\push{\ket{0}}&&\gate{H}&\gate{T}&\targ&\gate{T^\dagger}&\targ&\gate{T}&\targ&\gate{T^\dagger}&\gate{H}&\gate{S^\dagger}&\gate{H}&\gate{S^\dagger}&\gate{H}&\gate{T^\dagger}&\targ&\gate{T}&\targ&\gate{T^\dagger}&\targ&\gate{T}&\gate{H}&\meter&\cctrl{-3}&\cctrl{-1}&&
}}\end{array}.
\end{equation}
Using the fact that
\begin{equation*}
\begin{array}{c}\Qcircuit @C=.5em @R=0em @!R {
&\gate{T^\dagger}&\gate{H}&\gate{S^\dagger}&\gate{H}&\gate{S^\dagger}&\gate{H}&\gate{T^\dagger}&\qw&\push{=}&&\qw&\qw&\qw&\qw
}\end{array},
\end{equation*}
Circ.~(\ref{eq:cccz_deriv_7}) can be transformed into
\begin{equation*}
\begin{array}{c}\scalebox{.8}{\Qcircuit @C=.3em @R=0em @!R {
&&\qw&\qw&\qw&\qw&\qw&\qw&\qw&\qw&\qw&\qw&\ctrl{4}&\qw&\qw&\qw&\qw&\qw&\ctrl{1}&\qw&\qw&\qw\\
&&\qw&\qw&\qw&\qw&\qw&\qw&\qw&\qw&\ctrl{3}&\qw&\qw&\qw&\ctrl{3}&\qw&\qw&\qw&\controlo\qw&\qw&\qw&\qw\\
&&\qw&\qw&\qw&\qw&\qw&\ctrl{2}&\qw&\qw&\qw&\qw&\qw&\qw&\qw&\qw&\qw&\qw&\qw&\ctrl{1}&\qw&\qw\\
&&\qw&\qw&\qw&\ctrl{1}&\qw&\qw&\qw&\ctrl{1}&\qw&\qw&\qw&\qw&\qw&\qw&\qw&\qw&\qw&\control\qw&\qw&\qw\\
&\push{\ket{0}}&&\gate{H}&\gate{T}&\targ&\gate{T^\dagger}&\targ&\gate{T}&\targ&\targ&\gate{T}&\targ&\gate{T^\dagger}&\targ&\gate{T}&\gate{H}&\meter&\cctrl{-3}&\cctrl{-1}&&
}}\end{array}.
\end{equation*}
Thus, \cref{fig:comp_cccz}a is a decomposition of the CCCZ gate.
\end{proof}

\subsection{Fewer \texorpdfstring{$T$}{T}-depth decompositions}\label{sec:few_t_depth}

This subsection explains how to construct a decomposition for an $n$-controlled Pauli gate that keeps the currently known minimum $T$-count of $4n-6$, while achieving a smaller $T$-depth.

First, we noticed that a transformation similar to \cref{eq:rtof_1a,eq:rtof_1b} can be applied to a part of the temporary logical-AND gate~(\ref{eq:logical_and}), as shown below:
\begin{equation}\label{eq:rtof_2a}
\begin{array}{c}\scalebox{.8}{\Qcircuit @C=.3em @R=.2em @!R {
&\qw&\qw&\qw&\ctrl{2}&\targ&\gate{T^\dagger}&\targ&\qw&\qw&&&\ctrlo{1}&\qw&\ctrl{1}&\qw&\qw&&&\qw&\ctrl{1}&\qw&\ctrlo{1}&\qw\\
&\qw&\qw&\ctrl{1}&\qw&\targ&\gate{T^\dagger}&\targ&\qw&\qw&\push{=}&&\ctrlo{1}&\qw&\ctrl{1}&\qw&\qw&\push{=}&&\qw&\ctrl{1}&\qw&\ctrlo{1}&\qw\\
&\gate{H}&\gate{T}&\targ&\targ&\ctrl{-2}&\gate{T}&\ctrl{-2}&\gate{H}&\qw&&&\control\qw&\gate{S^\dagger}&\targ&\gate{S^\dagger}&\qw&&&\gate{S^\dagger}&\targ&\gate{S^\dagger}&\control\qw&\qw
}}\end{array}.
\end{equation}
Likewise, the following equation also holds:
\begin{equation}\label{eq:rtof_2b}
\begin{array}{c}\scalebox{.8}{\Qcircuit @C=.3em @R=.2em @!R {
&\qw&\targ&\gate{T^\dagger}&\targ&\ctrl{2}&\qw&\qw&\qw&\qw&&&\ctrlo{1}&\qw&\ctrl{1}&\qw&\qw&&&\qw&\ctrl{1}&\qw&\ctrlo{1}&\qw\\
&\qw&\targ&\gate{T^\dagger}&\targ&\qw&\ctrl{1}&\qw&\qw&\qw&\push{=}&&\ctrlo{1}&\qw&\ctrl{1}&\qw&\qw&\push{=}&&\qw&\ctrl{1}&\qw&\ctrlo{1}&\qw\\
&\gate{H}&\ctrl{-2}&\gate{T}&\ctrl{-2}&\targ&\targ&\gate{T}&\gate{H}&\qw&&&\control\qw&\gate{S^\dagger}&\targ&\gate{S^\dagger}&\qw&&&\gate{S^\dagger}&\targ&\gate{S^\dagger}&\control\qw&\qw
}}\end{array}.
\end{equation}

Using these, the three decompositions (Figs.~\ref{fig:comp_ccz}b, \ref{fig:comp_cccz}b, and \ref{fig:comp_cadd}b) can be derived. The derivation method is almost identical to that described in \cref{sec:few_cnot_count}, except that \cref{eq:rtof_2a,eq:rtof_2b} are used instead of \cref{eq:rtof_1a,eq:rtof_1b}, so it is omitted here.
\cref{fig:comp_ccz}b is a CCZ gate decomposition with $T$-count 4 and $T$-depth 2. However, this decomposition lacks novelty, as it can also be derived by sandwiching the CZ gate between a temporary logical-AND gate~(\ref{eq:logical_and}) and its uncomputation gate~(\ref{eq:uncomputation_logical_and}).
\Cref{fig:comp_cccz}b is a CCCZ gate decomposition with $T$-count 6 and $T$-depth 2. This decomposition successfully improves $T$-depth in the currently known minimal $T$-count decomposition~(\ref{eq:gidney_cccz}) which has $T$-depth 6.
\Cref{fig:comp_cadd}b shows a control qubit addition method. However, this also lacks novelty, as it can similarly be derived by sandwiching the controlled unitary gate between a temporary logical-AND gate~(\ref{eq:logical_and}) and its uncomputation gate~(\ref{eq:uncomputation_logical_and}).
While Figs.~\ref{fig:comp_ccz}b and \ref{fig:comp_cadd}b can both be derived using Gidney's method~\cite{gidney2018halving}, \cref{fig:comp_cccz}b can only be derived using our method.

In addition, although the number of auxiliary qubits increases, we have discovered a Clifford+$T$ gate implementation of an $n$-controlled Pauli gates with a smaller $T$-depth, while keeping the currently known minimum $T$-count of $4n-6$. We explain this below.

First, we noticed that the quantum circuit representing the double-controlled $-iX$ gate~(\ref{eq:ccix_selinger_minus}) can be rewritten as follows:
\begin{equation}\label{eq:rtof_3}
\begin{array}{c}\scalebox{.8}{\Qcircuit @C=.3em @R=.2em @!R {
&&\qw&\ctrl{3}&\targ&\qw&\gate{T^\dagger}&\qw&\targ&\ctrl{3}&\qw&\qw&&&\ctrl{1}&\qw&\ctrl{1}&\qw&\qw&&&\qw&\ctrl{1}&\qw&\ctrl{1}&\qw\\
&&\qw&\qw&\targ&\ctrl{2}&\gate{T^\dagger}&\ctrl{2}&\targ&\qw&\qw&\qw&\push{=}&&\ctrl{1}&\qw&\ctrl{1}&\qw&\qw&\push{=}&&\qw&\ctrl{1}&\qw&\ctrl{1}&\qw\\
&&\gate{H}&\qw&\ctrl{-2}&\qw&\gate{T}&\qw&\ctrl{-2}&\qw&\gate{H}&\qw&&&\control\qw&\gate{S}&\targ&\gate{S^\dagger}&\qw&&&\gate{S^\dagger}&\targ&\gate{S}&\control\qw&\qw\\
&\push{\ket{0}\hspace{-.8em}}&&\targ&\qw&\targ&\gate{T}&\targ&\qw&\targ&\qw&\push{\hspace{-.8em}\ket{0}}&&&&&&&&&&&&&&
}}\end{array}.
\end{equation}

Using these, the three decompositions (\cref{fig:comp_ccz}c, Circ.~(\ref{eq:cccz_useless}), and \cref{fig:comp_cadd}c) can be derived. The derivation method is almost identical to that described in \cref{sec:few_cnot_count}, except that \cref{eq:rtof_3} are used instead of \cref{eq:rtof_1a,eq:rtof_1b}, so it is omitted here.
\begin{equation}
\begin{array}{c}\scalebox{.8}{\Qcircuit @C=.3em @R=.2em @!R {&&\qw&\qw&\qw&\qw&\qw&\qw&\qw&\qw&\qw&\qw&\targ&\ctrl{5}&\gate{T^\dagger}&\ctrl{5}&\targ&\qw&\qw&\qw&\ctrl{1}&\qw&\qw&\qw\\
&&\qw&\qw&\qw&\qw&\qw&\qw&\qw&\qw&\qw&\ctrl{4}&\targ&\qw&\gate{T^\dagger}&\qw&\targ&\ctrl{4}&\qw&\qw&\control\qw&\qw&\qw&\qw\\
&&\qw&\qw&\ctrl{3}&\targ&\qw&\gate{T^\dagger}&\qw&\targ&\ctrl{3}&\qw&\qw&\qw&\qw&\qw&\qw&\qw&\qw&\qw&\qw&\ctrl{1}&\qw&\qw\\
&&\qw&\qw&\qw&\targ&\ctrl{2}&\gate{T^\dagger}&\ctrl{2}&\targ&\qw&\qw&\qw&\qw&\qw&\qw&\qw&\qw&\qw&\qw&\qw&\control\qw&\qw&\qw\\
&\push{\ket{0}}&&\gate{H}&\qw&\ctrl{-2}&\qw&\qw&\qw&\ctrl{-2}&\qw&\qw&\ctrl{-4}&\qw&\qw&\qw&\ctrl{-4}&\qw&\gate{H}&\meter&\cctrl{-3}&\cctrl{-1}&&\\
&&\push{\ket{0}\hspace{-.8em}}&&\targ&\qw&\targ&\gate{T}&\targ&\qw&\targ&\targ&\qw&\targ&\gate{T}&\targ&\qw&\targ&\qw&\push{\hspace{-2em}\ket{0}}&&&&
}}\end{array}\label{eq:cccz_useless}
\end{equation}
\Cref{fig:comp_ccz}c is a CCZ gate decomposition with $T$-count 4 and $T$-depth 1. However, this decomposition was already discovered by Jones~\cite{jones2013low}, so it lacks novelty.
Circuit~(\ref{eq:cccz_useless}) is a CCCZ gate decomposition. It has $T$-depth 2, just like \cref{fig:comp_cccz}b, but it requires four more CNOT gates and one additional auxiliary qubit compared to \cref{fig:comp_cccz}b. Therefore, Circ.~(\ref{eq:cccz_useless}) does not offer an advantage over \cref{fig:comp_cccz}b. 
\Cref{fig:comp_cadd}c shows a control qubit addition method. However, this decomposition was already discovered by Jones~\cite{jones2013low}, so it lacks novelty.

\section{Comparison}\label{sec:comp}

\begin{table*}
\centering
\begin{tabular}{ccccc}
\hline
&\# meas.&&&\\
&feedback&$T$-count&CNOT-count&$T$-depth\\
\hline \hline
Selinger~\cite{selinger2013quantum}&0&15&32&3\\
Jones~\cite{jones2013low}&2&8&17 or 19&1\\
Gidney~\cite{gidney2018halving}&2&8&13 or 15&2\\
Gidney and Jones~\cite{gidney2021cccz} (Circ.~(\ref{eq:gidney_cccz}))&1&\textbf{6}&8&6\\
Ours (\cref{fig:comp_cccz}a)&1&\textbf{6}&\textbf{6 or 8}&6\\
Ours (\cref{fig:comp_cccz}a')&1&\textbf{6}&\textbf{7}&6\\
Ours (\cref{fig:comp_cccz}b)&1&\textbf{6}&12 or 14&\textbf{2}\\
\hline
\end{tabular}
\caption{Comparison of computational resources between our proposed CCCZ gate decompositions and those from previous studies. The column name ``\# meas.\ feedback'' represents the number of measurement-based feedback controls. In the $T$-count column, the currently known minimal $T$-count of 6 is highlighted in bold. Bold values in the CNOT-count and $T$-depth columns indicate the smallest CNOT-count or $T$-depth among the decompositions with the known minimal $T$-count of 6. As shown in the table, for the known minimal $T$-count of 6, our proposed methods achieve a smaller CNOT-count or $T$-depth compared to previous research.}
\label{tb:comp_cccz}
\end{table*}

\begin{table*}
\centering
\begin{tabular}{cccc}
\hline
&\# meas.&&\\
&feedback&$T$-count&CNOT-count\\
\hline \hline
Selinger~\cite{selinger2013quantum}&$\pm$0&+8&+16\\
Ours (\cref{fig:comp_cadd}a)&+1&\textbf{+4}&\textbf{+3 or +4}\\
Gidney~\cite{gidney2018halving}, Ours (\cref{fig:comp_cadd}b)&+1&\textbf{+4}&+6 or +7\\
Jones~\cite{jones2013low}, Ours (\cref{fig:comp_cadd}c)&+1&\textbf{+4}&+8 or +9\\
\hline
\end{tabular}
\caption{Comparison of computational resources for control qubit addition methods. The column name ``\# meas.\ feedback'' represents the number of measurement-based feedback controls. The $T$-count and CNOT-count columns indicate how much the $T$-count or CNOT-count increases with each additional control qubit. The smallest increment in each case is highlighted in bold.}
\label{tb:comp_add_ctrl}
\end{table*}

\begin{table*}
\centering
\begin{tabular}{cccc}
\hline
&\# meas.&&CNOT-count\\
&feedback&$T$-count&(best, worst)\\
\hline \hline
Selinger~\cite{selinger2013quantum}, Ours (Sec.~\ref{sec:improve_selinger})&0&$8n-9$&$16n-16$\\
Jones~\cite{jones2013low}&$n-1$&$4n-4$&$8n-7$, $9n-8$\\
Gidney~\cite{gidney2018halving}&$n-1$&$4n-4$&$6n-5$, $7n-6$\\
Gidney and Jones~\cite{gidney2021cccz} (Circ.~(\ref{eq:gidney_cccz}) \& \cref{fig:comp_cadd}b)&$n-2$&$\bm{4n-6}$&$6n-10$, $7n-13$\\
Ours (Figs.~\ref{fig:comp_cccz}a \& \ref{fig:comp_cadd}a)&$n-2$&$\bm{4n-6}$&$\bm{3n-3}$\textbf{,} $\bm{4n-4}$\\
Ours (Figs.~\ref{fig:comp_cccz}a' \& \ref{fig:comp_cadd}a)&$n-2$&$\bm{4n-6}$&$\bm{3n-2}$\textbf{,} $\bm{4n-5}$\\
Ours (Figs.~\ref{fig:comp_cccz}a \& \ref{fig:comp_cadd}b)&$n-2$&$\bm{4n-6}$&$6n-12$, $7n-13$\\
Ours (Figs.~\ref{fig:comp_cccz}b \& \ref{fig:comp_cadd}b)&$n-2$&$\bm{4n-6}$&$6n-6$, $7n-7$\\
Ours (Figs.~\ref{fig:comp_cccz}b \& \ref{fig:comp_cadd}c)&$n-2$&$\bm{4n-6}$&$8n-12$, $9n-13$\\
\hline
\end{tabular}
\caption{Comparison of computational resources between our proposed $n$-controlled Pauli gate decompositions $(n \geq 3)$ and those from previous studies. When using measurement-based feedback control, the CNOT-count varies depending on the measurement outcome, so the CNOT-count column shows both the minimum and maximum values. Bold values in the $T$-count and CNOT-count columns indicate the smallest values, respectively. As shown in the table, we have discovered an $n$-controlled Pauli gate decomposition with a smaller CNOT-count than previous studies, while keeping the currently known minimal $T$-count of $4n - 6$.}
\label{tb:comp_cnz}
\end{table*}

The novel quantum gate decompositions that we have derived in this paper are the three CCCZ gate decompositions and the C$^n$Z gate decompositions that are created by adding control qubits to these CCCZ gate decompositions. In this section, we compare these with previous research and show their advantages.

\subsection{CCCZ gate decompositions}\label{sec:comp_cccz}

We compare our three novel CCCZ gate decompositions with previous research. \Cref{fig:comp_cccz} shows these decompositions. \Cref{tb:comp_cccz} presents a comparison of $T$-count, CNOT-count, and $T$-depth between our CCCZ gate decompositions and previous research.

First, we review the CCCZ gate decomposition (\ref{eq:gidney_cccz}) with the known minimal $T$-count, discovered by Gidney and Jones~\cite{gidney2021cccz}. This decomposition has $T$-count 6 and $T$-depth 6, with one measurement-based feedback control. The CNOT-count is 8 regardless of whether the measurement outcome is 0 or 1.

\Cref{fig:comp_cccz}a shows one of our CCCZ gate decompositions proposed in \cref{sec:few_cnot_count}. This decomposition also has $T$-count 6 and $T$-depth 6, with one measurement-based feedback control. The CNOT-count is 6 when the measurement outcome is 0, and 8 when it is 1. When the measurement result is 0, this decomposition reduces the CNOT-count by 2 compared to Circ.~(\ref{eq:gidney_cccz}). Moreover, \cref{fig:comp_cccz}a is visually simpler than Circ.~(\ref{eq:gidney_cccz}).

\Cref{fig:comp_cccz}a' shows another decomposition proposed in \cref{sec:few_cnot_count}. Unlike \cref{fig:comp_cccz}a, this decomposition features a constant CNOT-count regardless of the measurement outcome. It also has $T$-count 6 and $T$-depth 6 with one measurement-based feedback control. The CNOT-count is 7, which is consistently 1 less than Circ.~(\ref{eq:gidney_cccz}).

Finally, \cref{fig:comp_cccz}b presents our CCCZ gate decomposition proposed in \cref{sec:few_t_depth}. This decomposition was designed to minimize $T$-depth while keeping $T$-count 6. It also involves one measurement-based feedback control, and its $T$-depth is reduced to 2, which is 4 less than Circ.~(\ref{eq:gidney_cccz}).

\subsection{Control qubit addition methods}\label{sec:comp_add_ctrl}

We now compare methods for adding control qubits to any controlled unitary gate. Among the three methods we derived in \cref{sec:main_decompositions}, two are identical to those found in previous research. \Cref{fig:comp_cadd} shows a visual comparison of these methods, and \cref{tb:comp_add_ctrl} compares the increase in $T$-count and CNOT-count for each method when adding a control qubit.

\Cref{fig:comp_cadd}a shows our method, derived in \cref{sec:few_cnot_count}. When this method is used to add a control qubit, $T$-count increases by 4, and the number of measurement-based feedback control increases by 1. The CNOT-count increases by 4 when the measurement outcome is 0 and by 3 when it is 1. \Cref{fig:comp_cadd}a can be viewed as a generalization of the Toffoli gate decomposition by Paler \textit{et al.}~\cite{paler2022realistic}.

\Cref{fig:comp_cadd}b presents a method discovered by Gidney~\cite{gidney2018halving}, which we also derived in \cref{sec:few_t_depth}. When using this method, $T$-count also increases by 4 per added a control qubit, and the number of measurement-based feedback control increases by 1. The CNOT-count increases by 6 when the measurement outcome is 0 and by 7 when it is 1.

Finally, \cref{fig:comp_cadd}c shows a method by Jones~\cite{jones2013low}, which we also derived in \cref{sec:few_t_depth}. When adding a control qubit with this method, $T$-count increases by 4, and the number of measurement-based feedback control increases by 1. The CNOT-count increases by 8 when the measurement outcome is 0 and by 9 when it is 1.

\subsection{\texorpdfstring{C$^n$Z}{CnZ} gate decompositions}\label{sec:comp_cnz}

\begin{figure}
    \centering
    \includegraphics[width=\linewidth]{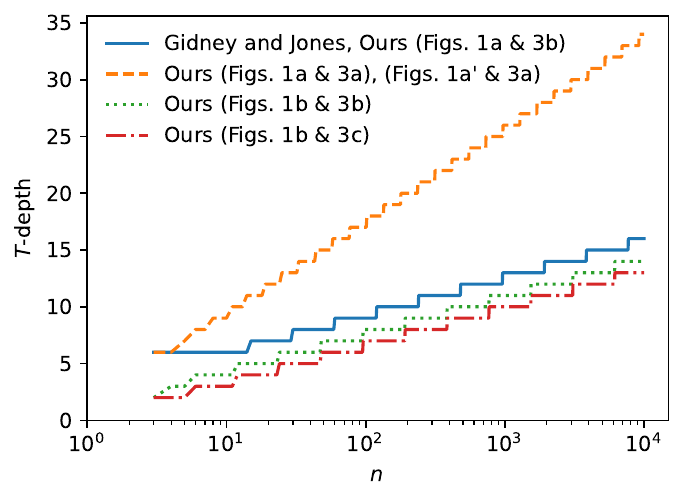}
    \caption{Comparison of $T$-depth between our proposed decompositions and previous studies for $n$-controlled Pauli gate decompositions $(n \geq 3)$ with $T$-count $4n - 6$. As shown in the graph, we have discovered an $n$-controlled Pauli gate decomposition with smaller $T$-depth than previous studies, while keeping the currently known minimal $T$-count of $4n - 6$.}
    \label{fig:comp_cnz_tdepth}
\end{figure}

Finally, we compare C$^n$Z gate decompositions for $n \geq 3$. C$^n$Z gate decompositions are constructed by repeatedly applying the methods for adding control qubits to CZ, CCZ, or CCCZ gates. \Cref{tb:comp_cnz} compares the $T$-count and CNOT-count of these methods, and \cref{fig:comp_cnz_tdepth} compares the minimal $T$-depth that can be achieved using the method with the  known minimal $T$-count of $4n-6$.

Before detailing each methods, we explain how the minimal $T$-depth is derived. When using any method from \cref{fig:comp_cadd} to add a control qubit to a C$^k$Z gate, one of the qubits in the C$^k$Z gate must be chosen as the control qubit of the single-controlled $U$ gates in \cref{fig:comp_cadd}. Since there are $k+1$ choices for this, we select the option that minimizes the $T$-depth. By repeatedly applying this method, we derive the minimal $T$-depth for each C$^n$Z gate decomposition, as shown in \cref{fig:comp_cnz_tdepth}.

Gidney and Jones~\cite{gidney2021cccz} constructed a C$^n$Z gate decomposition with the smallest known $T$-count of $4n-6$ by applying their CCCZ gate decomposition~(\ref{eq:gidney_cccz}) and the control qubit addition method of \cref{fig:comp_cadd}b $n-3$ times. This decomposition requires $n-2$ measurement-based feedback controls, and the CNOT-count varies between $6n-10$ and $7n-13$, depending on the measurement outcomes. The $T$-depth, which was not specified in their paper, was derived and shown in \cref{fig:comp_cnz_tdepth}.

Similarly, we constructed several C$^n$Z gate decompositions by applying the control qubit addition methods (\cref{fig:comp_cadd}) to our CCCZ gate decompositions (\cref{fig:comp_cccz}). These decompositions keep the minimal $T$-count $4n-6$ while achieving lower CNOT-count or $T$-depth, as shown in \cref{tb:comp_cnz} and \cref{fig:comp_cnz_tdepth}.

In particular, the decomposition constructed from \cref{fig:comp_cccz}a and \cref{fig:comp_cadd}a achieves a CNOT-count between $3n-3$ and $4n-4$. The decomposition constructed from \cref{fig:comp_cccz}a' and \cref{fig:comp_cadd}a has a CNOT-count between $3n-2$ and $4n-5$. Both are smaller than the CNOT-count of the decomposition by Gidney and Jones~\cite{gidney2021cccz}.

As shown in \cref{fig:comp_cnz_tdepth}, the decompositions constructed from \cref{fig:comp_cccz}b and \cref{fig:comp_cadd}b, as well as \cref{fig:comp_cccz}b and \cref{fig:comp_cadd}c, achieve smaller $T$-depth while keeping the known minimal $T$-count of $4n-6$.

\section{Conclusion}\label{sec:conclusion}
This paper focuses on the multi-controlled Pauli gate decompositions with an eye to both current non-fault-tolerant and future fault-tolerant quantum computers.

The contribution of this research is that we have discovered a method for systematically constructing multi-controlled Pauli gate decompositions with smaller CNOT-count or $T$-depth while keeping the same $T$-count as the currently known minimum $T$-count decomposition. For example, we discovered that the CCCZ gate decomposition can be constructed with CNOT-count 7 or $T$-depth 2 while keeping $T$-count 6, the smallest $T$-count known to date. Since multi-controlled Pauli gates are used in the quantum circuits of various quantum algorithms, the discovery of these efficient decompositions will lead to a reduction in the execution time of quantum circuits. This is expected to improve the computational efficiency of various quantum algorithms. This will be an important step towards the future practical application of quantum computers. The discovery is also made possible by the realization of an important procedure that is key to systematically constructing multi-controlled Pauli gate decompositions. This procedure not only deepens our theoretical understanding of quantum circuit decompositions, but also has the potential to lead to the discovery of more efficient gate decompositions that have yet to be discovered.

Future challenges include improving the efficiency of other important multi-qubit operations beyond multi-controlled Pauli gates in quantum circuits. In addition, verifying the effectiveness of efficient decompositions of multi-controlled Pauli gates through implementation on actual quantum devices will be essential.

\begin{acknowledgments}
    This calculation has been done using NVIDIA GPGPU at the Institute for Physics of Intelligence ($i\pi$), School of Science, the University of Tokyo.
    KMN is supported by the Daikin Endowed Research Unit: ``Research on Physics of Intelligence'', School of Science, the University of Tokyo.
    ST acknowledges support by the Endowed Project for Quantum Software Research and Education, The University of Tokyo (\url{https://qsw.phys.s.u-tokyo.ac.jp/}) and the Center of Innovations for Sustainable Quantum AI (\url{https://sqai.jp/}) (JST Grant Number JPMJPF2221).
\end{acknowledgments}

\bibliography{main}

\onecolumngrid
\clearpage
\appendix

\section{Derivation of Eq.~(\ref{eq:cnz_base})}
Let the initial state of $(m+n)$-qubits be $\ket{\psi}$ as follows:
\begin{equation}\label{eq:cnot_h_cnot_with_barrier}
\begin{array}{c}\scalebox{1}{\Qcircuit @C=1em @R=.2em @!R {
&&&0&&1&&2&&3&&4&&5&\\
\lstick{}&{/^m}\qw&\qw\barrier[0em]{2}&\qw&\qw\barrier[0em]{2}&\qw&\gate{U}\barrier[0em]{2}&\qw&\qw\barrier[0em]{2}&\qw&\qw\barrier[0em]{2}&\qw&\qw\barrier[0em]{2}&\qw&\qw\\
\lstick{}&{/^n}\qw&\qw&\qw&\ctrl{1}&\qw&\qw&\qw&\qw&\qw&\qw&\qw&\control\qw&\qw&\qw
\inputgroupv{2}{3}{1em}{.95em}{\ket{\psi}}\\
&&\lstick{\ket{0}}&\qw&\targ&\qw&\ctrl{-2}&\qw&\gate{H}&\qw&\meter&\cw&\cctrl{-1}&&
}}\end{array}.
\end{equation}
The state $\ket{\psi}$ can be expressed using $m$-qubit quantum state $\ket{\phi_1}$ and $n$-qubit quantum state $\ket{\phi_2}$ as follows:
\begin{equation}
\ket{\psi} = a_{00}\ket{\phi_1}\ket{\phi_2} + a_{01}\ket{\phi_1}\ket{\bm{1}_2} + a_{10}\ket{\bm{1}_1}\ket{\phi_2} + a_{11}\ket{\bm{1}_1}\ket{\bm{1}_2},
\end{equation}
where $\ket{\bm{1}_1}$ denotes all of $m$-qubit states are $\ket{1}$, $\ket{\bm{1}_2}$ denotes all of $n$-qubit states are $\ket{1}$, $\braket{\phi_1|\bm{1}_1}=0$, $\braket{\phi_2|\bm{1}_2}=0$.

In this case, the state at position 0 in the Circ.~(\ref{eq:cnot_h_cnot_with_barrier}) can be expressed as follows:
\begin{equation}
a_{00}\ket{\phi_1}\ket{\phi_2}\ket{0} + a_{01}\ket{\phi_1}\ket{\bm{1}_2}\ket{0} + a_{10}\ket{\bm{1}_1}\ket{\phi_2}\ket{0} + a_{11}\ket{\bm{1}_1}\ket{\bm{1}_2}\ket{0}.
\end{equation}
The multi-controlled NOT gate between positions 0 and 1 in Circ.~(\ref{eq:cnot_h_cnot_with_barrier}) transforms $\ket{\phi_2}\ket{0}$ into $\ket{\phi_2}\ket{0}$ and $\ket{\bm{1}_2}\ket{0}$ into $\ket{\bm{1}_2}\ket{1}$. Therefore, the state at position 1 in Circ.~(\ref{eq:cnot_h_cnot_with_barrier}) can be expressed as follows:
\begin{equation}
a_{00}\ket{\phi_1}\ket{\phi_2}\ket{0} + a_{01}\ket{\phi_1}\ket{\bm{1}_2}\ket{1} + a_{10}\ket{\bm{1}_1}\ket{\phi_2}\ket{0} + a_{11}\ket{\bm{1}_1}\ket{\bm{1}_2}\ket{1}.
\end{equation}
Applying the controlled $U$ gate, the state at position 2 in Circ.~(\ref{eq:cnot_h_cnot_with_barrier}) can be expressed as follows:
\begin{equation}
a_{00}\ket{\phi_1}\ket{\phi_2}\ket{0} + a_{01}(U\ket{\phi_1})\ket{\bm{1}_2}\ket{1} + a_{10}\ket{\bm{1}_1}\ket{\phi_2}\ket{0} + a_{11}(U\ket{\bm{1}_1})\ket{\bm{1}_2}\ket{1}.
\end{equation}
Applying the Hadamard gate, the state at position 3 in Circ.~(\ref{eq:cnot_h_cnot_with_barrier}) can be expressed as follows:
\begin{align}
& \frac{a_{00}}{\sqrt{2}}\ket{\phi_1}\ket{\phi_2}\ket{0}
+ \frac{a_{00}}{\sqrt{2}}\ket{\phi_1}\ket{\phi_2}\ket{1}
+ \frac{a_{01}}{\sqrt{2}}(U\ket{\phi_1})\ket{\bm{1}_2}\ket{0}
- \frac{a_{01}}{\sqrt{2}}(U\ket{\phi_1})\ket{\bm{1}_2}\ket{1}\notag\\
&+\frac{a_{10}}{\sqrt{2}}\ket{\bm{1}_1}\ket{\phi_2}\ket{0}
+ \frac{a_{10}}{\sqrt{2}}\ket{\bm{1}_1}\ket{\phi_2}\ket{1}
+ \frac{a_{11}}{\sqrt{2}}(U\ket{\bm{1}_1})\ket{\bm{1}_2}\ket{0}
- \frac{a_{11}}{\sqrt{2}}(U\ket{\bm{1}_1})\ket{\bm{1}_2}\ket{1}
\end{align}
Then, the state at position 4 in Circ.~(\ref{eq:cnot_h_cnot_with_barrier}) can be expressed as follows:
\begin{itemize}
    \item if measurement result is 0:
    \begin{equation}
        a_{00}\ket{\phi_1}\ket{\phi_2} + a_{01}(U\ket{\phi_1})\ket{\bm{1}_2} + a_{10}\ket{\bm{1}_1}\ket{\phi_2} + a_{11}(U\ket{\bm{1}_1})\ket{\bm{1}_2}
    \end{equation}
    \item if measurement result is 1:
    \begin{equation}
        a_{00}\ket{\phi_1}\ket{\phi_2} - a_{01}(U\ket{\phi_1})\ket{\bm{1}_2} + a_{10}\ket{\bm{1}_1}\ket{\phi_2} - a_{11}(U\ket{\bm{1}_1})\ket{\bm{1}_2}
    \end{equation}
\end{itemize}
Therefore, the state at position 5 in Circ.~(\ref{eq:cnot_h_cnot_with_barrier}) can be expressed as follows:
\begin{equation}
a_{00}\ket{\phi_1}\ket{\phi_2} + a_{01}(U\ket{\phi_1})\ket{\bm{1}_2} + a_{10}\ket{\bm{1}_1}\ket{\phi_2} + a_{11}(U\ket{\bm{1}_1})\ket{\bm{1}_2}
\end{equation}
This state is equal to the state which applied $n$-controlled $U$ gate to $\ket{\psi}$.

\section{Derivation of Fig.~\ref{fig:comp_cccz}a'}\label{apx:cccz_min_cnot_count_alternative}

The decomposition \cref{fig:comp_cccz}a has CNOT-count 6 or 8, but it is also possible to create a decomposition with CNOT-count 7. The details of this decomposition and a broad outline of its derivation are discussed below. 

First, a slight modification of \cref{eq:cnz_base} results in the following equation:
\begin{equation}\label{eq:similar_cnz_base}
\begin{array}{c}\scalebox{1}{\Qcircuit @C=.6em @R=0em @!R {
&{/^m}\qw&\qw&\qw&\qw&\qw&\ctrl{2}&\qw&\qw&\qw&\qw&&&{/^m}\qw&\qw&\ctrl{1}&\qw&\qw\\
&{/^n}\qw&\qw&\ctrl{1}&\qw&\qw&\qw&\qw&\control\qw&\qw&\qw&\push{\raisebox{1em}{=}}&&{/^n}\qw&\qw&\control\qw&\qw&\qw\\
&&\lstick{\ket{0}}&\targ&\gate{H}&\gate{X}&\targ&\meter&\cctrlo{-1}&&&&&&&&&
}}\end{array}
\end{equation}
By transforming it in the same manner as the previous proof and substituting \cref{eq:rtof_1a} twice, Fig.~\ref{fig:comp_cccz}a' is obtained.
The decomposition described in \cref{fig:comp_cccz}a' achieves CNOT-count 7.

\end{document}